\documentclass[twocolumn]{aastex631}
\usepackage{soul}

\usepackage{CJK}
\usepackage{amsmath} 
\usepackage{amsbsy}
\usepackage{enumitem}

\newcommand\nsample{441}

\shorttitle{Color Gradients of Galaxies at $z > 4$}
\shortauthors{Jin, Ho \& Sun}

\graphicspath{{./}{figures/}}

\begin{document}
\begin{CJK*}{UTF8}{gbsn}

\title{A High Incidence of Central Star Formation Inferred from the Color Gradients of Galaxies at $z>4$}

\correspondingauthor{Wen Sun}
\email{sunwen@stu.pku.edu.cn}

\author[0009-0003-4862-2925]{Bingcheng Jin (金秉诚)}
\affiliation{Kavli Institute for Astronomy and Astrophysics, Peking University, Beijing 100871, China}
\affiliation{Department of Astronomy, School of Physics, Peking University, Beijing 100871, China}

\author[0000-0001-6947-5846]{Luis C. Ho}
\affiliation{Kavli Institute for Astronomy and Astrophysics, Peking University, Beijing 100871, China}
\affiliation{Department of Astronomy, School of Physics, Peking University, Beijing 100871, China}

\author[0000-0003-3995-4859]{Wen Sun (孙文)}
\affiliation{Kavli Institute for Astronomy and Astrophysics, Peking University, Beijing 100871, China}
\affiliation{Department of Astronomy, School of Physics, Peking University, Beijing 100871, China}

\begin{abstract}
We study the rest-frame ultraviolet-optical color gradients of \nsample galaxies at $4<z<8$ by characterizing the wavelength dependence of their structural parameters derived from simultaneously fitting the seven-band NIRCam images acquired with the James Webb Space Telescope. Distinct from trends observed at lower redshifts, where most galaxies exhibit negative color gradients whereby galaxy centers are redder than their outskirts, in high-redshift galaxies positive color gradients are just as common as or even outnumber negative color gradients. Varying stellar population, dust, and active galactic nuclei can contribute to the observed color gradient. We show that for the majority of our sample, the observed color gradients principally reflect radial variations in stellar population, without strong contribution from dust reddening or contamination from active galactic nuclei. The sign and magnitude of the color profile depend systematically on the global properties of the galaxy: positive color gradients, characteristic of centrally concentrated star formation or outside-in growth, are found preferentially in galaxies of lower stellar mass, smaller size, and bluer spectral energy distribution.
\end{abstract}


\section{Introduction} \label{sec:intro}

The structure and morphology of galaxies, and how these properties change with time, reflect how galaxies evolve through their varied accretion and merger histories \citep{whiteGalaxyFormationHierarchical1991, steinmetz2002a}. Observations across a wide range of redshifts increasingly suggest that galaxies experience ``inside-out'' growth \citep[e.g.,][]{vandokkumGrowthMassiveGalaxies2010, perezEvolutionGalaxiesResolved2013, marianColorGradientsReflect2018, frankelInsideoutGrowthGalactic2019}. After an early phase in which a central bulge rapidly builds up, an extended disk gradually accumulates through accretion and minor mergers. This scenario is supported by a variety of observational clues, ranging from spatially resolved, coherent star formation \citep{nelsonWhereStarsForm2016,matharuFirstLookSpatially2024}, evidence of inside-out quenching \citep{tacchellaEvidenceMatureBulges2015, abdurroufEvolutionSpatiallyResolved2018}, and the redshift evolution of galaxy size \citep{vanderwel3DHSTCANDELSEvolution2014, suessHalfmassRadii0002019}.

Among the many possible empirical indicators that can be used to diagnose the complex evolutionary pathway of a galaxy, one of the most direct and arguably simplest approaches is through the measurement of its radial color gradient. The radial gradient of color in a galaxy reflects the spatial variation of stellar age, metallicity, and dust attenuation. However, the interplay among these factors is complex and varies strongly with galaxy type, environment, and cosmic epoch. For example, the color gradients of nearby early-type galaxies largely reflect radial variations in metallicity induced by their metal-rich centers \citep{vaderThreeColorSurface1988, marianColorGradientsReflect2018}, while in late-type galaxies negative color gradients arise from younger stellar populations in their outer region \citep{ryderRelationshipPresentStar1994}. Systematic studies of galaxies in the local Universe (e.g., \citealt{munoz-mateosSpecificStarFormation2007, kelvinGalaxyMassAssembly2012}) find a preponderance of negative color gradients---color profiles that are redder in the center and bluer in the outskirts---with a propensity for stronger negative color gradients in galaxies of late-type or lower \cite{Sersic1968} indices \citep{kennedyGalaxyMassAssembly2015}. Negative color gradients, albeit of weaker strength and larger scatter, continue to be seen in galaxies out to $z \approx 1-2$ \citep{suessHalfmassRadii0002019, millerColorGradientsHalfmass2023a, vanderwelStellarHalfMassRadii2024}. 

Dust attenuation presents a major practical challenge to the interpretation of observed color gradients, for the effects of reddening can be confused with intrinsic variations in stellar population \citep{grahamInclinationDustcorrectedGalaxy2008, millerEarlyJWSTImaging2022, zhangDustAttenuationDust2023}. \citet{Wang2017MNRAS} found that the radial color profiles of $0.4 < z < 1.4$ star-forming main sequence galaxies are affected more by dust than stellar population variations. Similarly, the larger galaxy sizes measured in the rest-frame ultraviolet (UV) compared to the optical, often interpreted as evidence that the central regions of galaxies have older stellar populations (e.g., \citealt{moslehEvolutionMassSizeRelation2012}), may arise, instead, from heavier dust attenuation in the centers of massive galaxies \citep{nedkovaUVCANDELSRoleDust2024}. However, the role of dust attenuation is less clear at high redshifts. Dust may be less of a concern at $z\gtrsim 4$ \citep[e.g.,][]{fudamoto2020, Looser2025, mitsuhashiALMACRISTALSurveyWidespread2024} given the limited time for metal enrichment and the lower dust-to-gas ratios expected, especially for low-mass systems \citep{liDusttogasDusttometalRatio2019}.

The James Webb Space Telescope (JWST; \citealt{Gardner2006}) ushers in an unprecedented opportunity to resolve fainter, redder, and more distant galaxies, especially in their rest-frame optical and near-infrared wavelengths \citep[e.g.,][]{abdurroufSpatiallyResolvedStellar2023,gimenez-arteagaSpatiallyResolvedProperties2023a,matharuFirstLookSpatially2023}. Inside-out growth may have begun very early, judging by the core-disk structure of the galaxy at $z=7.43$ reported by \citet{bakerInsideoutGrowthEarly2024}. Studying the size evolution of galaxies at $0.5 < z < 5$, \citet{jiJADESRestframeUVtoNIR2024} suggest that negative color gradients have already been established in massive quiescent systems at $z > 3$. By contrast, \citet{onoCensusRestframeOptical2024} find that galaxies at $z>4$ have almost identical effective radii in the rest-frame UV and optical, indicating little evidence for color gradients, which, if anything, are more likely positive instead of negative \citep{morishitaEnhancedSubkpcScale2024}. At even higher redshifts, no trends between morphology and colors have been detected in galaxies at $z=7-12$ \citep{yangEarlyResultsGLASSJWST2022, treuEarlyResultsGLASSJWST2023}. The apparent discrepancies in these results likely stem from differences in the sample selection of galaxy types, underscoring the need for a more systematic and thorough investigation of galaxy color gradients at high redshift. It remains to be seen to what extent robust color gradients have already been established at these early epochs, and, if so, what the main driver behind them is and what galaxy properties correlate with them.

We take advantage of JWST images taken with the Near Infrared Camera \citep[NIRCam;][]{2023PASP..135b8001R} from the Cosmic Evolution Early Release Science (CEERS; \citealt{finkelsteinCEERSKeyPaper2023, Finkelstein2025}) program to conduct a systematic study of the color gradients of \nsample~galaxies at $z>4$ and investigate their dependence on galaxy properties. The CEERS survey is designed to study the processes of galaxy assembly and the abundance and physical nature of galaxies in the early Universe. Although broadly similar topics have been studied recently with JWST, as mentioned above, our approach differs in several key aspects. Measuring structural parameters with robust uncertainties for high-redshift galaxies remains challenging, notwithstanding the impressive data from JWST. We obtain accurate parametric fits of the galaxies by simultaneously modeling the seven-band NIRCam images. Detailed mock experiments are performed to account for systematic uncertainties in the final error budget. Section~\ref{sec:obs} describes the observational material and sample definition. Section~\ref{sec:methods} introduces our image analysis technique and methodology to quantity color gradient. Results are given in Section~\ref{sec:results}, their implications are discussed in Section~\ref{sec:discuss}, and Section~\ref{sec:summary} summarizes the main findings. We assume a cosmology with $H_0 = 67.7$ km s$^{-1}$ Mpc$^{-1}$, $\Omega_m = 0.307$, and $\Omega_{\Lambda} = 0.693$ \citep{planckcollaborationPlanck2018Results2020}. All magnitudes are expressed in the absolute bolometric system of \cite{Oke1983}.

\section{Data and Sample Definition} \label{sec:obs}

We use the NIRCam imaging data from the CEERS program, which covers $\sim100$ arcmin$^2$ with three short-wavelength bands (F115W, F150W, F200W) and four long-wavelength bands (F277W, F356W, F410M, F444W). The mosaics of the 10 pointings overlap with the CANDELS Extended Groth Strip (EGS) region, one of the most well-studied fields \citep{koekemoerCANDELSCosmicAssembly2011} in the era of the Hubble Space Telescope (HST). The CEERS team has published their data release after two epochs of observations \citep{finkelsteinCEERSKeyPaper2023}. We directly download the latest version of the reduced data from the CEERS website\footnote{\url{https://ceers.github.io/releases.html}}. The data reduction process is discussed thoroughly in \citet{bagleyCEERSEpochNIRCam2023}, and the reduction scripts are available on GitHub\footnote{\url{https://github.com/ceers/ceers-nircam}}. Although various approaches have been adopted to reduce these data \citep[e.g.,][]{morishitaEnhancedSubkpcScale2024,zhuangAGNsHostGalaxies2024}, each tailored for a particular purpose, the structural parameter measurements are mostly consistent with each other (e.g., see Figure 9 in \citealt{sunStructureMorphologyGalaxies2023}).

The point-spread function (PSF) is crucial for determining accurate structural parameters, especially for compact, high-redshift galaxies. For the four pointings of the first epoch of the CEERS observations (CEERS1, CEERS2, CEERS3, and CEERS6), \citet[][see their Section~3.2 and Appendix~A]{sunStructureMorphologyGalaxies2023} constructed a PSF model for each band by stacking isolated, unsaturated point-like sources that have the spectral energy distribution (SED) of stars. Because of the relatively small number of objects available, they did not produce a separate PSF for each pointing, but, instead, for each filter, the objects in all pointings were combined to create a master, stacked PSF of high signal-to-noise ratio. PSF variations across different pointings of the same epoch should be small because the pointings were taken close in time and utilize the same dither pattern \citep{bagleyCEERSEpochNIRCam2023}. A $81 \times 81$ pixel cutout was extracted for each object, and all the cutouts are 4 times oversampled to align their centers after removing the background. The final PSF was constructed by mean-combining the individual stars and resampling the oversampled images to their original resolution. For the pointings from the first epoch, we directly adopt the PSF library created by \cite{sunStructureMorphologyGalaxies2023}, and we exactly follow their procedure to create the PSFs for the remaining six pointings from the second epoch observations.

To prepare for the image analysis, we carry out the following steps to extract sources from the CEERS mosaics. We apply \texttt{Photutils} \citep{bradleyAstropyPhotutils122024} to perform $1\,\sigma$ source detection on an image of high signal-to-noise ratio created by stacking the images of all the long-wavelength bands (F277W, F356W, F410M, F444W), weighted by their error. Many high-redshift galaxies are faint. If they are outshined by bright neighboring sources, we use the \texttt{deblend\char`_sources} function to separate them. The catalog is then constructed with the deblended segmentation map. We crossmatch our source catalog with the CANDELS photometric redshift catalog \citep{stefanonCANDELSMultiwavelengthCatalogs2017}, whose detection was based on the HST F160W band, with a $5\,\sigma$ limiting magnitude of 27.6. Our study focuses on galaxies in the redshift range $4<z<8$. We adopt Stefanon et al.'s catalog for convenience because it provides both photometric redshifts and self-consistent photometry across 22 bands, covering wavelengths from 0.4 to 8 $\mu$m, which will benefit the inference of physical properties used for our analysis (Section~\ref{subsec:dependency}).

For sources exposed in more than one telescope pointing, which often appear on the edge of the mosaics, we exclude redundant cutouts and ensure that the surviving cutout has as many available pixels as possible for background estimation and model fitting. After removing galaxies that do not have complete imaging, we obtain an initial sample of 669 galaxies (Table~\ref{tab:format}).

\begin{figure*}
\begin{center}
\includegraphics[width=0.65\textwidth]{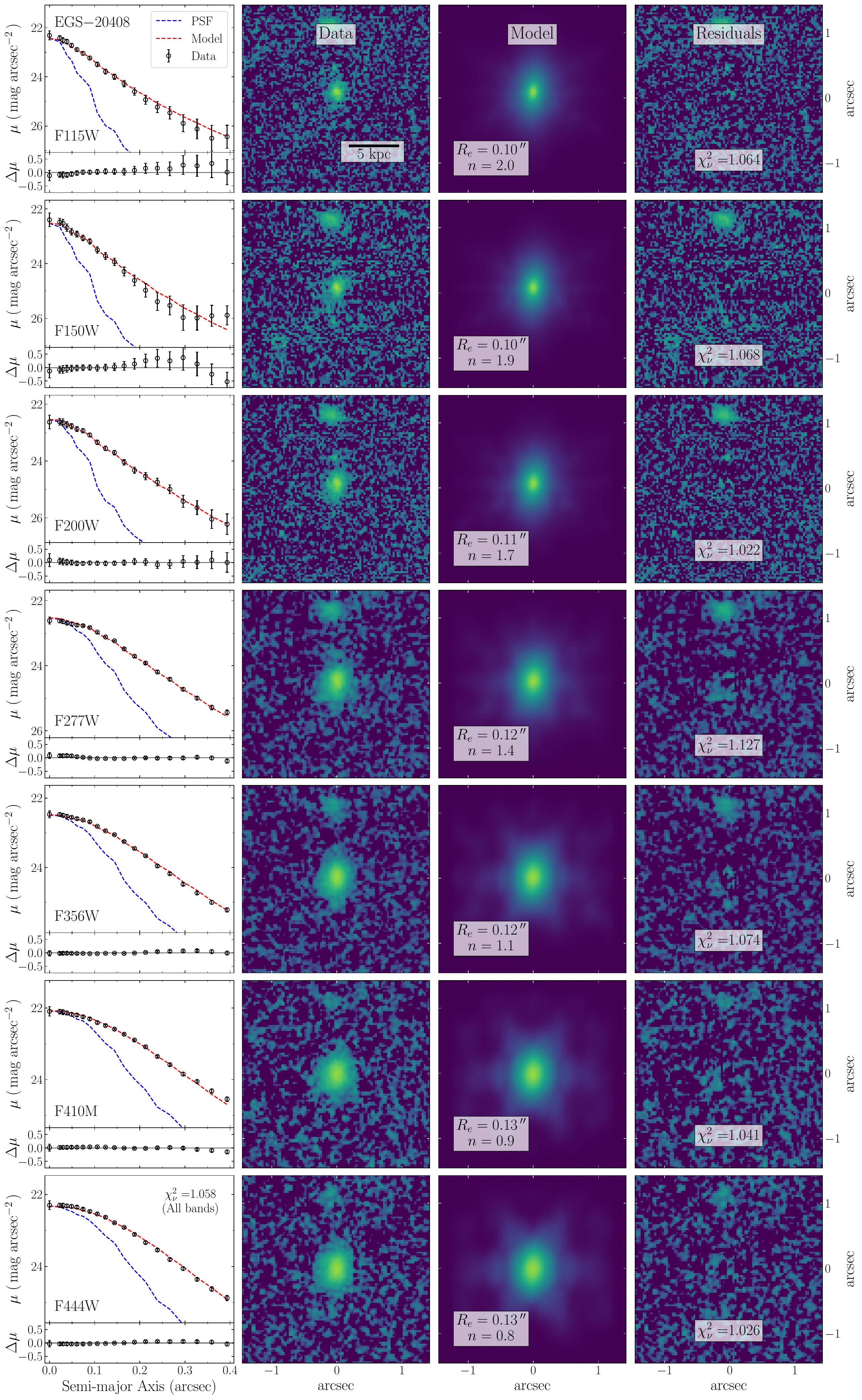}
\end{center}
\caption{Simultaneous, multiband fitting of the galaxy EGS$-$20408 \citep{stefanonCANDELSMultiwavelengthCatalogs2017} at $z=5.373$. Rows from top to bottom are the results for the F115W, F150W, F200W, F277W, F356W, F410M, and F444W bands. Columns from left to right are the surface brightness profile, input science image, model image (convolved with the PSF), and residual image. The $\chi^2_\nu$ for each band is given in the residual image, and the structural parameters are shown in the lower-left corner of the model image. In the surface brightness profile, the black points with error bars are measured by aperture photometry, the red dashed line shows the model profile, and the blue dashed line represents the PSF, scaled to match the model profile. \label{fig:singlecomp}
}
\end{figure*}

\section{Image Analysis} \label{sec:methods}

\subsection{Model Fitting} \label{subsec:galfitm}

We describe the galaxy surface brightness distribution by the S\'ersic function,

\begin{equation}
\mu(R) = \mu_e \exp\left\{-\kappa\left[\left(\frac{R}{R_e}\right)^{1/n} - 1\right]\right\},
\label{eq:sersic}
\end{equation}

\noindent 
where $R_e$ is the effective radius of the galaxy that contains half of the total flux, $\mu_e$ is the surface brightness at $R_e$, S\'ersic index $n$ specifies the shape of the profile, and $\kappa$ is related to $n$ by the incomplete gamma function, $\Gamma\left(2n\right)=2\gamma\left(2n,\kappa\right)$ \citep{Ciotti1991}. The light distribution of the galaxy is modeled using \texttt{GalfitM} \citep{hausslerMegaMorphMultiwavelengthMeasurement2013,vikaMegaMorphMultiwavelengthMeasurement2013,vikaMegaMorphClassifyingGalaxy2015}, a modified version of \texttt{Galfit} \citep{pengDetailedStructuralDecomposition2002, pengDETAILEDDECOMPOSITIONGALAXY2010} that can fit simultaneously images in multiple bands. Model parameters across different filters are constrained with a Chebyshev polynomial function whose order the user can control. Previous studies show that constraining $n$ and $R_e$ using a polynomial improves the accuracy and stability of the multi-wavelength model, especially in situations where some bands may suffer from marginal signal-to-noise ratio \citep{vikaMegaMorphMultiwavelengthMeasurement2013}.

For subsequent analysis, we generate image cutouts of size 7 times the \cite{kronPhotometryCompleteSample1980} radius, a size that contains sufficient pixels for background estimation while still computationally manageable. We use \texttt{source\char`_catalog} to measure basic non-parametric quantities, such as the Kron flux, axis ratio ($q$), and position angle ($\Theta$), which serve as initial guesses for the parametric fits. Following \citet{sunStructureMorphologyGalaxies2023}, contaminating sources in the cutout are masked by dilating the segmentation map with a ``growth radius'' (\citealt{hoCarnegieIrvineGalaxySurvey2011,huangCarnegieIrvineGalaxySurvey2013}; see Section~\ref{sec:obs}). We do not mask contaminating sources that are adjacent to the target galaxy, as they may significantly impact the light distribution of the main galaxy; instead, we simultaneously fit them using multiple components (see \citealt{Liyang+2023}). We estimate and then subtract the mean value of the local background by excluding all detections, including the target itself, in the cutout with dilating masks. To produce the weight (sigma) image, we scale the error map from the \texttt{ERR} extension such that its level equals the background standard deviation of the science image.

Unlike previous applications of \texttt{GalfitM} (e.g., \citealt{vikaMegaMorphClassifyingGalaxy2015,sunStructureMorphologyGalaxies2023}), in this study we apply additional constraints to ensure that the parameters $n$ and $R_e$ change monotonically with wavelength, in accord with the observed behavior of galaxies, whose wavelength-dependent radial variation in light profile manifests as color gradients, which are induced by systematic changes in the radial distribution of stellar population and dust reddening (e.g., \citealt{kelvinGalaxyMassAssembly2012}). While invoking a higher-order polynomial to link the structural parameters across different filters may yield a mathematically superior fit, the solution may not be physically meaningful. Hence, we only permit $n$ and $R_e$ to vary linearly with wavelength. We verified that our main conclusions are insensitive to this choice. The convergence of the fit can be quite sensitive to the initial values of the free parameters. We set the initial value of $n$ to 2, close to the median value found by \cite{sunStructureMorphologyGalaxies2023}, and $R_e$ to the median value of the half-light radius measured using \texttt{Photutils} \citep{bradleyAstropyPhotutils122024}. Initially set to the Kron aperture flux, the magnitudes are free to vary with wavelength following a Chebyshev polynomials of order 7. The initial guesses for $q$ and $\Theta$ are available from the source detection procedure, and during the fit, these parameters are kept constant with wavelength.

\begin{figure*}[t!]
\includegraphics[width=0.98\textwidth]{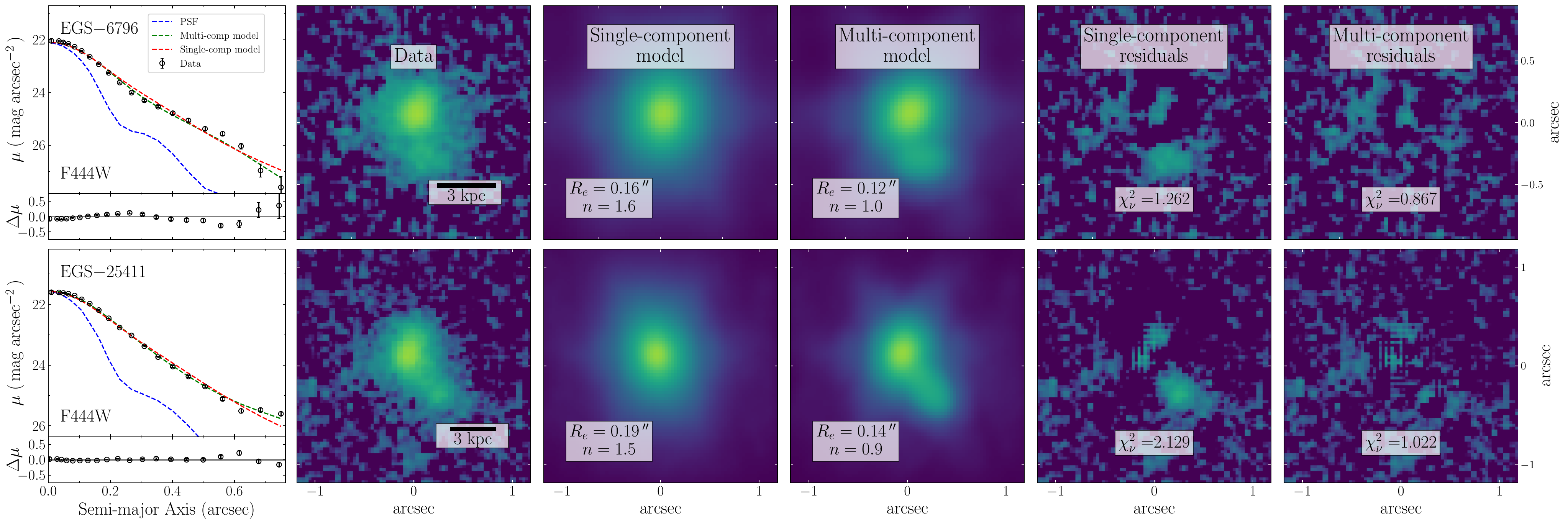}
\caption{Treatment of blended sources, to avoid biased measurement of structural parameters contaminated by neighboring sources. In the surface brightness profile, the black points with error bars are measured by aperture photometry, the red dashed line shows the single-component model profile, the green dashed line the multi-component model, which includes the target galaxy and the contaminating neighbor, and the blue dashed line represents the PSF, scaled to match the model profile. It is clear that the outer region of the target galaxy is heavily contaminated by the neighbor, and the additional component is necessary to measure the structural parameters of the target galaxy accurately. The images show, from left to right, the original image, the single-component model, the multi-component model, and the respective residuals for the two models. \label{fig:multicomp}}
\end{figure*}

\begin{figure}
\begin{center}
\includegraphics[width=0.44\textwidth]{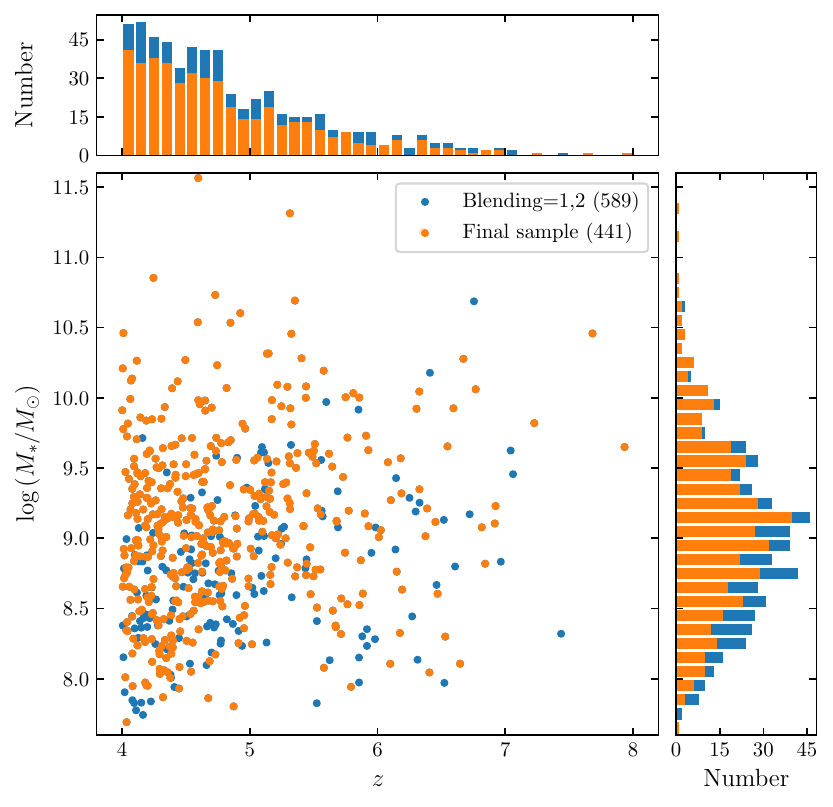}
\end{center}
\caption{Distribution of redshift and stellar mass of our final sample of 441 galaxies (orange points). Blue points correspond to galaxies with blending flag 1 or 2 in the initial sample. The stellar mass spans a wide range, from $M_* = 10^{7.5}$ to $10^{11}\, M_\odot$. The extreme massive galaxies with $M_* > 10^{11} \,M_\odot$ at $z \gtrsim 4$ should be treated with caution; they are likely to be low-redshift interlopers introduced by errors in their photometric redshifts.
\label{fig:sampdist}}
\end{figure}

Blended sources, when not properly masked, can lead to spurious results \citep{onoCensusRestframeOptical2024}. To guard against this effect, we visually inspect every cutout and its associated segmentation map to assign a flag on its degree of blending: (1) isolated, (2) blended by adjacent objects that are properly segmented, and (3) heavily blended by a contaminating source. For isolated targets (flag~1), the fitting process is straightforward, and a single S{\'e}rsic component suffices to describe the light distribution (Figure~\ref{fig:singlecomp}). Additional components are needed for flag~2 sources (Figure~\ref{fig:multicomp}), and these more complex fits take longer to converge. The following analysis only makes use of flag~1 and flag~2 sources (Figure~\ref{fig:sampdist}), and we do not consider the 80 galaxies with flag~3 classification. 

\subsection{Uncertainties of the Structural Parameters} \label{subsec:uncertainty}

As with \texttt{Galfit}, \texttt{GalfitM} estimates the uncertainties using the curvature of the $\chi^2$ surface around the best-fit solution. Many studies have shown that the uncertainties returned by \texttt{Galfit} or \texttt{GalfitM} are underestimated \citep{hausslerGALAPAGOS2GalfitMGAMA2022,nedkovaBulgeDiscDecomposition2024}. To address this issue, our error estimation is based on mock simulations similar to those described in \citet{zhuangStarformingMainSequence2022}, where we generate simulated images with known input parameters to resemble real observations. After creating a clean image for each galaxy with its best-fit parameters and convolving it with the PSF model, we mimic realistic observations by adding the Poisson noise of the source and considering the background noise level in each specific image. We generate 100 mocks for each galaxy and fit them with \texttt{GalfitM} using the same configuration applied to the actual observations. We adopt the median value of these input-output experiments as the final measurement and incorporate the standard deviation into the final error budget. 

For simplicity, the mock images are based on single-component S\'ersic models. We caution that the real uncertainties may still be underestimated because of possible irregular features or substructures within the galaxy. However, our analysis depends only on robust measurements of the integrated magnitude and global structural parameters $n$ and $R_e$ at different wavelengths, which are not strongly influenced by the presence of substructures \citep[e.g.,][]{pengDETAILEDDECOMPOSITIONGALAXY2010}. We do not expect that the effects of irregular features and substructures will substantially alter our estimation of uncertainties. Although flag~2 sources comprise less than $6$\% of our final sample, we caution that their parameter uncertainties are likely underestimated, as source blending can introduce additional measurement errors. However, given their small proportion, this effect should not significantly impact our overall results.

As discussed in \citet{sunStructureMorphologyGalaxies2023}, the measurements are biased by cosmological effects: the magnitudes of the faintest galaxies at $z=4$ are underestimated by 0.3~mag, and the sizes are also underestimated, especially for small galaxies. The scatter of the measurements is more significant for galaxies with large S\'ersic indices, being most conspicuous at the highest redshifts. We closely follow the mock analysis described in Appendix~B of \citet{sunStructureMorphologyGalaxies2023} to correct the measurement biases and incorporate the uncertainties into the final error budget.

\subsection{Quantification of Color Gradient} \label{subsec:proxy}

A variety of strategies have been employed to study the color gradients of high-redshift galaxies (e.g., \citealt{millerColorGradientsHalfmass2023a, jiJADESRestframeUVtoNIR2024, vanderwelStellarHalfMassRadii2024}). We focus on three simple measures that can be derived straightforwardly from our analysis. Following \cite{vulcaniGalaxyMassAssembly2014b}, we define the wavelength variation of the S{\'e}rsic index as

\begin{equation}
\mathcal{N} = \frac{n^{b}}{n^{r}},
\label{eq:nvar}
\end{equation}

\noindent where $n^{b}$ and $n^{r}$ denote the S{\'e}rsic index measured in the blue and red band, respectively. For example, an early-type galaxy with a redder, more bulge-dominated central region would have a larger value of $n$ at longer wavelengths, resulting in $\mathcal{N}<1$. In the same spirit, the wavelength variation of the effective radius can be expressed as

\begin{equation}
\mathcal{R} = \frac{R_e^{b}}{R_e^{r}},
\label{eq:rvar}
\end{equation}

\noindent with $R_e^{b}$ and $R_e^{r}$ the effective radius measured in the blue and red band, respectively. A galaxy with an outskirt bluer than its center would have a larger $R_e$ at shorter wavelengths, leading to $\mathcal{R}>1$. In short, negative color gradients (color profiles that become redder toward the center) are characterized by $\mathcal{N}<1$ and $\mathcal{R}>1$, whereas positive color gradients (color profiles that become bluer toward the center) are indicated by $\mathcal{N}>1$ and $\mathcal{R}<1$. Additionally, we adopt \citep{marianColorGradientsReflect2018}

\begin{equation}
\nabla = \frac{\Delta (\mu^{b}(R)-\mu^{r}(R))}{\Delta \log R},
\label{eq:nabla}
\end{equation}

\noindent where $\mu^{b}(R)$ and $\mu^{r}(R)$ are the respective surface brightness profiles as a function of radius $R$. For the CEERS data in hand, we find that the radial range $R = (0.1-1.2)\,R_e$ effectively captures the color gradient over an area with reliable signal-to-noise ratios. The quantity $\nabla$ directly measures the color gradient, while $\mathcal{R}$ and $\mathcal{N}$ are indirect proxies for the color gradient that enjoy the privilege of being more robust and less sensitive to artificial effects introduced by the definition of color gradient.

Table~\ref{tab:format} summarizes the three measures of color gradient for the sample. We adopt the optical band closest to rest-frame 5000~\AA\ as the red band, and for the blue we choose the band nearest to rest-frame 2000~\AA\ to represent the UV to trace star-forming activity. The rest-frame wavelength range $2000-5000$~\AA\ is fully covered by the NIRCam observations for galaxies in our redshift range of interest ($4<z<8$).

To estimate the true uncertainties of the color gradient measurements and to ascertain the limits to which such measurements can be made, we conduct realistic mock simulations similar to those described in Section~\ref{subsec:uncertainty}. The mocks are designed to study the influence of galaxy magnitude and structural parameters, the most critical of which is galaxy size, considering that the analysis hinges on the degree to which we can measure the multi-band internal structure of the systems. The input-output experiments, described in Appendix~\ref{app:mock}, show that $\mathcal{R}$ becomes highly uncertain and biased for galaxies with $R_e \lesssim 0\farcs04$. The measurements for $\mathcal{N}$ and $\nabla$ are even more demanding, starting to break down when $R_e \lesssim 0\farcs06$. We therefore apply a size cut of $R_e=0\farcs06$ to all the bands, a reasonable criterion in view of the PSF sizes of the NIRCam bands ($\rm FWHM \approx 0\farcs06-0\farcs16$; \citealt{rigbySciencePerformanceJWST2023, zhuangCharacterizationJWSTNIRCam2024}). After some experimentation, we find that a magnitude limit of $\rm F150W \approx 28$ also needs to be imposed. In total, \nsample~galaxies with no or only moderately serious levels of blending (blending flag 1 or 2; see Section~\ref{subsec:galfitm}) satisfy our size and magnitude cut (Figure~\ref{fig:sampdist}).

\begin{figure}
\centering
\includegraphics[width=0.45\textwidth]{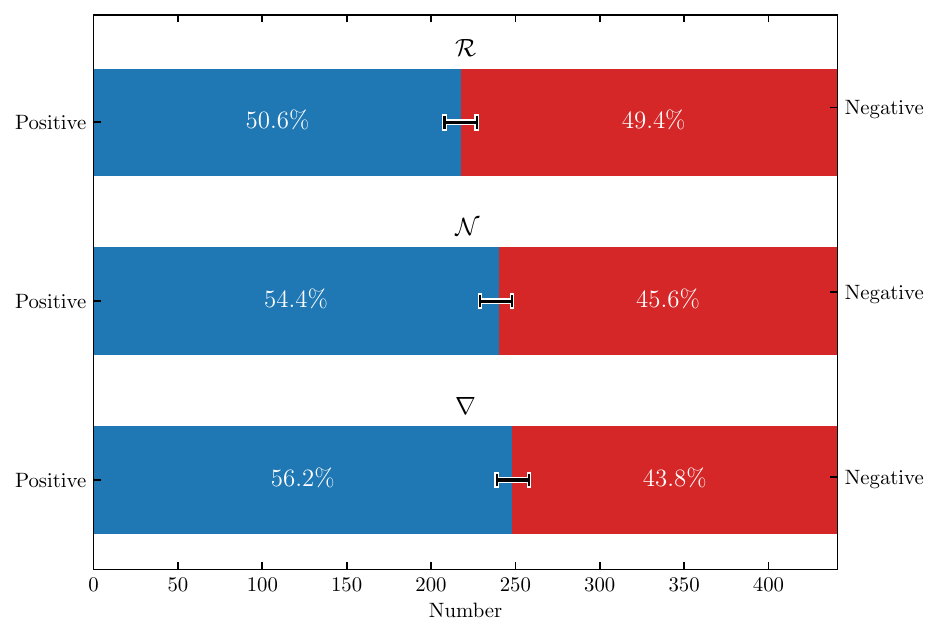}
\caption{Statistics of color gradients for the \nsample~galaxies in our sample, given separately for the three methods ($\mathcal{R}$, $\mathcal{N}$, $\nabla$) used to measure color gradient.
\label{fig:statsbar}}
\end{figure} 

\begin{figure*} [t]
\gridline{\fig{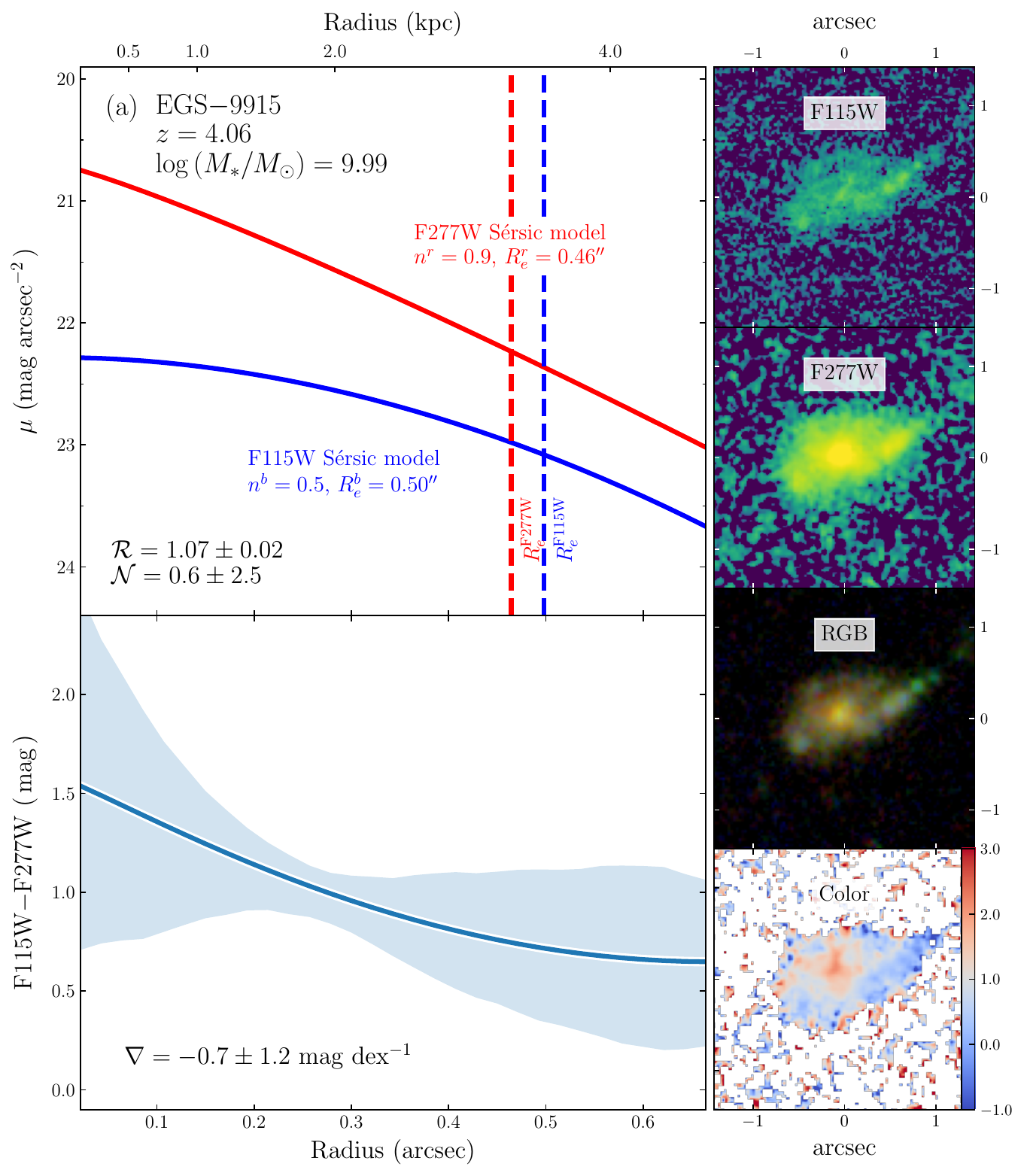}{0.45\textwidth}{}
\fig{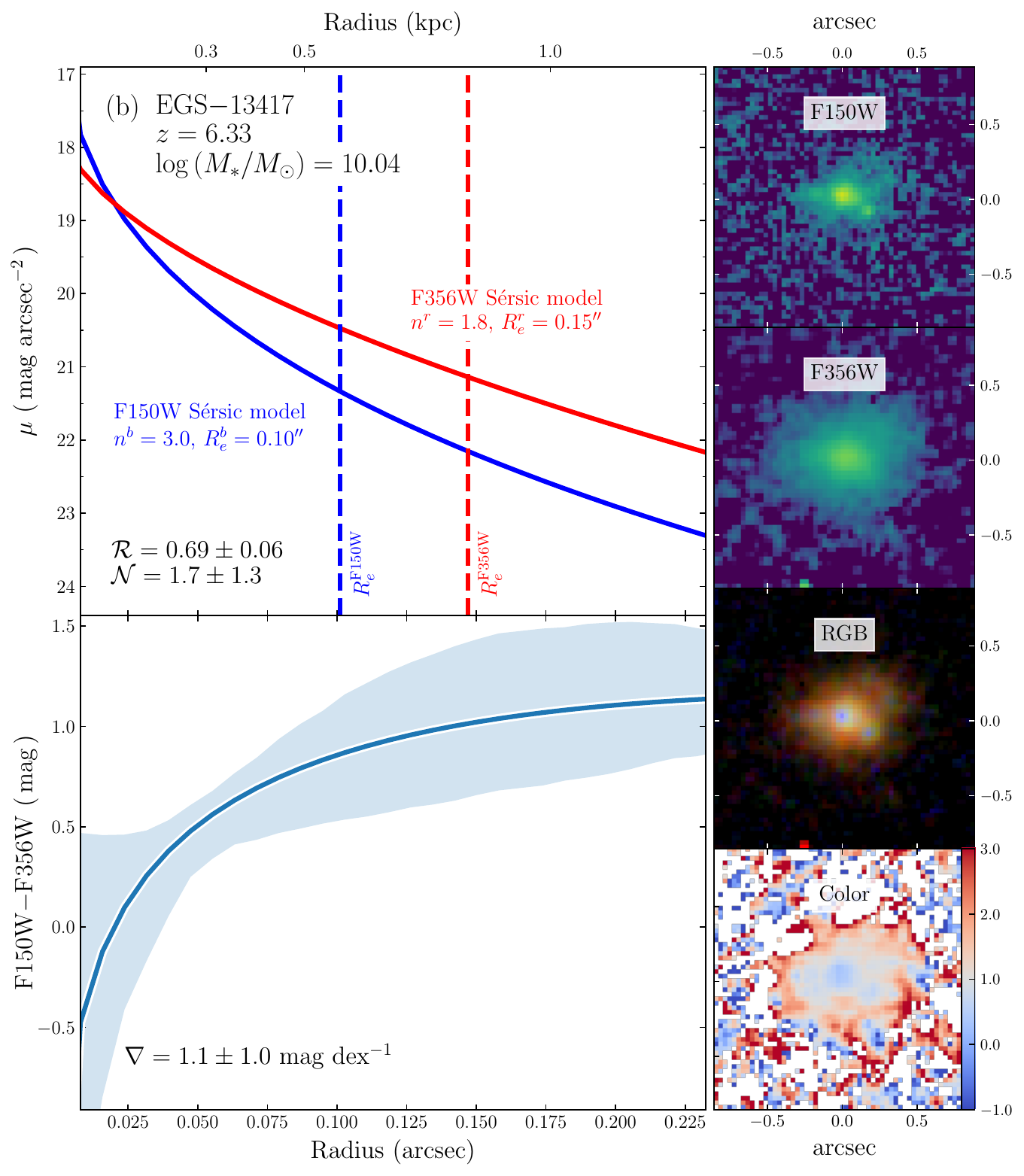}{0.45\textwidth}{}
}
\caption{Example of galaxies with (a) negative and (b) positive color gradient in our sample. In each panel, the left column shows the best-fit S{\'e}rsic model profile for the rest-frame UV and optical bands (top) and the UV$-$optical color profile (bottom), where the blue shaded area shows the 16th and 84th percentile of the color profile derived by sampling the structural parameter according to its uncertainty. The right column of each panel shows the two-dimensional visualizations of the color gradients: from top to bottom, the panels show the science images from the rest-frame UV and optical, the pseudo-color image (red = F444W, green = F277W, blue = F150W), and the PSF-matched UV$-$optical color image. \label{fig:colorprof}
}
\end{figure*}

\section{Results} \label{sec:results}

\subsection{Overall Statistics of Color Gradients at $z > 4$} \label{subsec:stats}

Galaxies generally exhibit color gradients. At low redshifts, most of the gradients are negative, in the sense that the color tends to be redder toward the center \citep[e.g.,][]{Zaritsky+1994, vulcaniGalaxyMassAssembly2014b}. Positive color gradients are rare. Merely 10\% of galaxies at $z \lesssim 1$ possess positive metallicity gradients (e.g., \citealt{Perez-Montero+2016} for $M_*=10^9-10^{11}\,M_\odot$ and \citealt{Carton+2018} for 
$M_*=10^7-10^{11}\,M_\odot$), and even at $1<z<3$ the incidence rises only to $\sim 20\%$ (\citealt{Simons+2021} and \citealt{Tissera+2022} for 
$M_* = 10^{8.5}-10^{11}\,M_\odot$; \citealt{SunXunda+2025} for $M_*>10^9\,M_\odot$). At the higher redshifts ($z > 4$) probed by our study, most galaxies also have well-established color gradients, but there is a distinctively different balance: a significantly larger proportion of galaxies have positive instead of negative color gradients. Among our final sample of \nsample~galaxies that are amenable to analysis, the majority ($\sim 50\%-60\%$) have positive color gradients. Negative color gradients are much less common at high redshifts. This may be due to distinct physical conditions in our higher redshift range, or perhaps the lower stellar masses (median $M_* = 10^{9.1}\, M_{\odot}$) of the galaxies in our sample. Although there are minor differences in the statistics depending on the method used to measure color gradient ($\mathcal{R}$, $\mathcal{N}$, or $\nabla$), the overall trend is robust (Figure~\ref{fig:statsbar}). 

Figure~\ref{fig:colorprof} illustrates the two situations encountered in our sample. The left panel shows a typical example of a galaxy with a negative color gradient, EGS$-$9915 at $z = 4.06$, whose color becomes bluer with increasing radius ($\nabla = -0.7\pm{1.0}~\rm mag\,dex^{-1}$).  The galaxy has a larger $R_e$ (0\farcs50 vs. 0\farcs46) and smaller $n$ (0.5 vs. 0.9) in the UV compared to the optical, indicating that it possesses a compact, redder central region and a blue, presumably star-forming extended component. In contrast, the galaxy EGS$-$13417 at $z = 6.33$ shown on the right panel behaves in the opposite manner: $R_e$ increases (0\farcs10 vs. 0\farcs15) while $n$ decreases (3.0 vs. 1.8) toward longer wavelength, leading to a positive color gradient ($\nabla = 1.1\pm1.0~\rm mag\,dex^{-1}$). The red core of EGS$-$9915 and the blue center of EGS$-$13417 are obvious from visual inspection of the RGB image formed by combining F444W, F277W, and F150W. The color gradient is also discernible from the {UV$-$optical} color image, which was created by matching the PSF of the two bands using \texttt{create\_matching\_kernel} in \texttt{photutils} with the kernel SplitCosineBellWindow.

\begin{figure}
\centering
\includegraphics[width=0.95\linewidth]{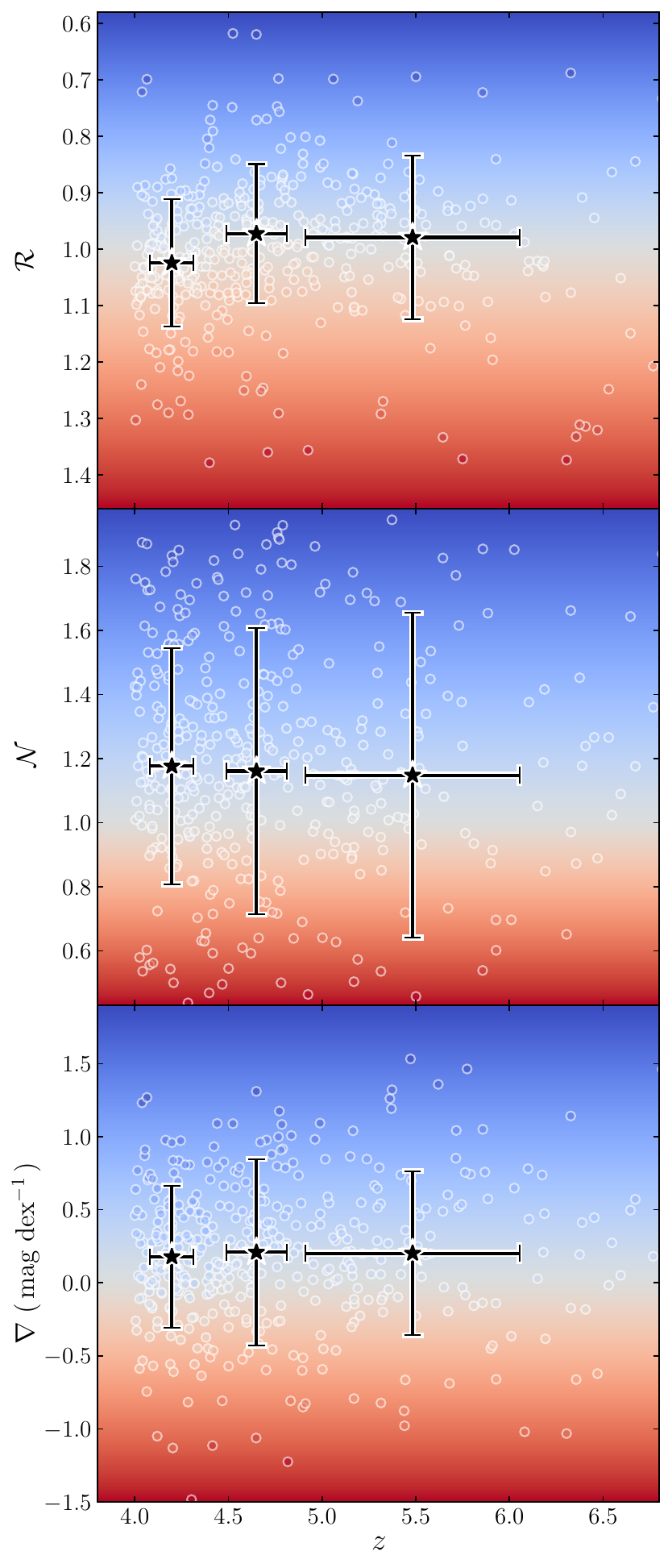}
\caption{The redshift evolution of the color gradient, quantified using the variation of the effective radius $\mathcal{R}$ (Equation~\ref{eq:rvar}; top), variation of the S\'ersic index $\mathcal{N}$ (Equation~\ref{eq:nvar}; middle), and the color gradient $\nabla$ (Equation~\ref{eq:nabla}; bottom). The $x$-axis stands for the median redshift of each redshift bin. The open white points correspond to individual measurements. The black stars indicate the median value for each redshift bin, with the error bars giving the $16$th and $84$th percentile of the distribution. The background gradient describes the region where the color gradient is positive (blue) or negative (red). \label{fig:zevo}}
\end{figure}

Figure~\ref{fig:zevo} examines whether the incidence of color gradients and the relative proportion of negative versus positive color gradients vary with redshift within the range ($4<z<8$) probed here. Because the measurement error increases toward high redshift, we bin the galaxies based on the median error of the measurement, dividing the sample into redshift bins of $4\leq z<4.4$, $4.4\leq z<5.0$, and $5.0\leq z <8$. The median error of each bin is $2-3$ times smaller than the scatter, and the sample is divided almost equally (Table~\ref{tab:zevo}). Galaxies with both negative and positive gradients are clearly common at high redshifts, but the relative incidence of each type remains unchanged within this redshift interval, no matter which of the color gradient measures is considered (Table~\ref{tab:zevo}).

\begin{deluxetable*}{cccccc}
    \tabletypesize{\scriptsize}
    \tablewidth{0pt}
    \tablecaption{Redshift Evolution of Color Gradient\label{tab:zevo}}
    \tablehead{
        \colhead{Redshift Range} & \colhead{Median Redshift} & \colhead{Number} & \colhead{$\mathcal{R}$} & \colhead{$\mathcal{N}$} &\colhead{$\nabla$ (mag$\,$arcsec$^{-1}$)}\\
        \colhead{(1)} & \colhead{(2)} & \colhead{(3)} & \colhead{(4)} & \colhead{(5)} & \colhead{(6)}
    }
    \startdata
        $4\leq z < 4.4$ & 4.20 & 151 & $1.02^{+0.10}_{-0.07}$ & $1.30^{+0.38}_{-0.36}$ & $0.28^{+0.39}_{-0.43}$\\ [0.5 em]
        $4.4\leq z < 5.0$ & 4.67 & 152 & $0.96^{+0.09}_{-0.10}$ & $1.29^{+0.51}_{-0.47}$ & $0.29^{+0.55}_{-0.59}$\\ [0.5 em]
        $5.0\leq z < 8.0$  & 5.58 & 138 & $1.00^{+0.09}_{-0.10}$ & $1.38^{+0.42}_{-0.65}$ & $0.50^{+0.62}_{-0.32}$\\  [0.5 em]
    \enddata
    \tablecomments{
    Col. (1): Redshift range.
    Col. (2): Median redshift.
    Col. (3): Number of galaxies.
    Cols. (4)--(6): Median value of $\mathcal{R}$, $\mathcal{N}$, and $\nabla$ and the $16$th and $84$th percent of their distributions.
    }
\end{deluxetable*}

\begin{figure*}
\centering
\includegraphics[width=0.9\textwidth]{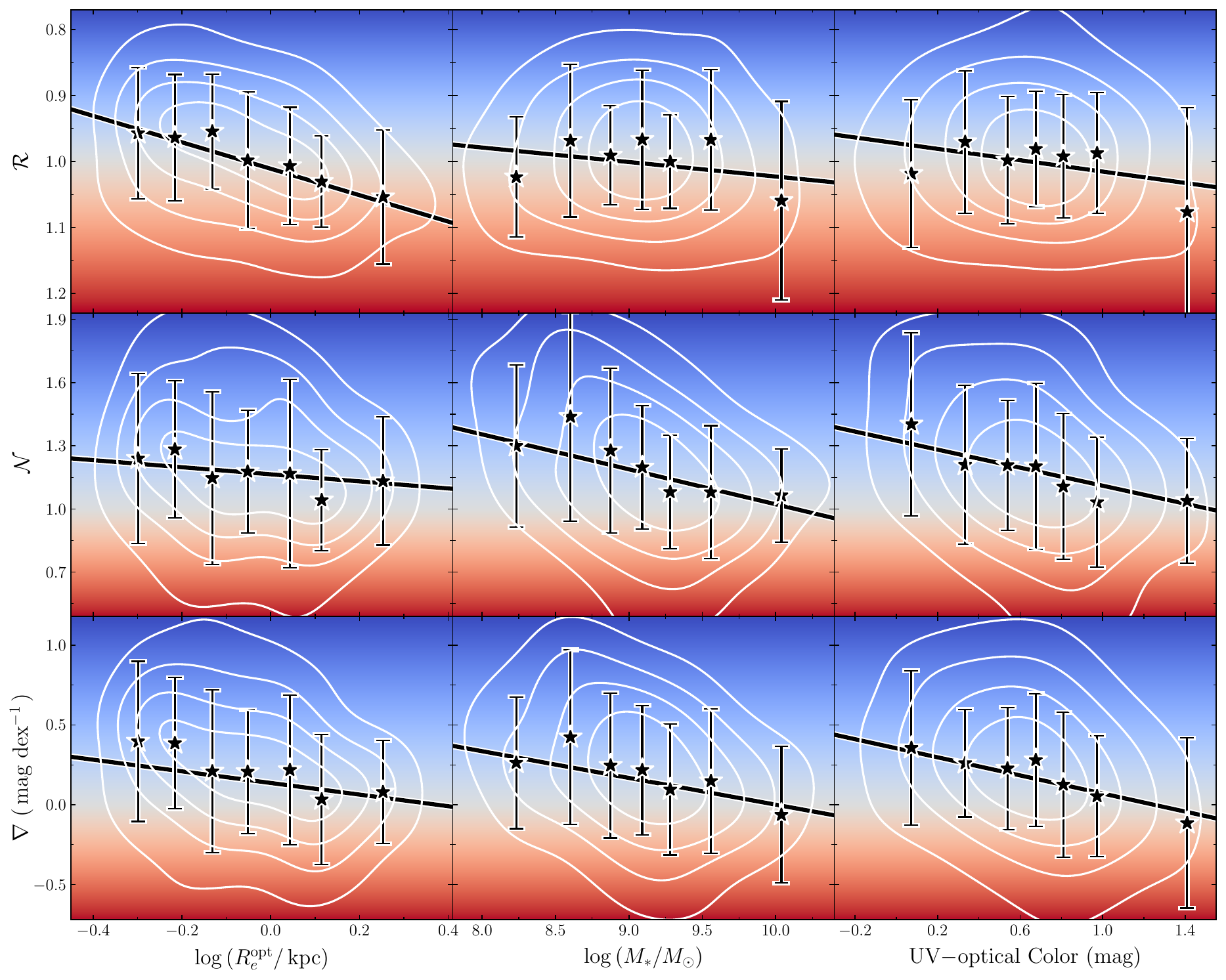}
\caption{Dependence on galaxy properties of the color gradient, quantified using the variation of the effective radius $\mathcal{R}$ (Equation~\ref{eq:rvar}, top), variation of the S\'ersic index $\mathcal{N}$ (Equation~\ref{eq:nvar}, middle), and the color gradient $\nabla$ (Equation~\ref{eq:nabla}, bottom). From left to right, the columns show the dependence on galaxy optical effective radius ($R_e^{\rm opt}$), stellar masses ($M_*$), and the rest-frame UV-to-optical color. The white contours denote the density distribution of the data points, the stars show the running median of the data binned so that each bin contains at least 60 galaxies, and the error bars indicate the standard deviation. The solid line gives the best-fit linear regression of the individual measurements. The background gradient denotes the region where the color gradient is positive (blue) or negative (red). \label{fig:dep}}
\end{figure*}

\begin{deluxetable}{ccrcl}
    \tabletypesize{\scriptsize}
    \tablewidth{0pt}
    \tablecaption{Best-fit Results of Linear Regression \label{tab:fitresult}}
    \tablehead{
        \colhead{Label} &  \colhead{$X$} & \colhead{$s$} & \colhead{$X_0$} & \colhead{$p$-value} \\
        \colhead{(1)} &  \colhead{(2)} & \colhead{(3)} & \colhead{(4)} & \colhead{(5)}
    }
    \startdata
        \multicolumn{5}{c}{$\mathcal{R}$ $= s(X-X_0) + 1$}\\ [1.5ex]
        (a) & $\log\, (R_e^{\rm opt}/\rm kpc)$ & $0.20\pm0.03$ & $-0.06\pm0.03$ & $0.000$*\\
        (b) & $\log\, (M_\star/M_{\odot})$ & $0.02\pm0.01$ & $8.96\pm0.29$ & $0.027$*\\
        (c) & UV$-$optical color (mag) & $0.04\pm0.01$ & $0.64\pm0.14$ & $0.000$*\\
        (d) & $\log\,(\rm sSFR/yr^{-1})$ & $-0.04\pm0.02$ & $-8.27\pm0.18$ & $0.042$* \\
        (e) & $q$ & $0.08\pm0.06$ & $0.45\pm0.08$ & $0.163$\\
        (f) & $\log\, n^{\rm opt}$ & $0.07\pm0.08$ & $-0.17\pm0.43$ & $0.420$\\
        \hline
        \multicolumn{5}{c}{$\mathcal{N}= s(X-X_0) $ $+ 1$}\\ [1.5ex]
        (g) & $\log\, (R_e^{\rm opt}/\rm kpc)$ & $-0.17\pm0.11$ & $0.98\pm0.68$ & $0.130$\\
        (h) & $\log\, (M_\star/M_{\odot})$ & $-0.17\pm0.03$ & $10.13\pm0.24$ & $0.000$*\\
        (i) & UV$-$optical color (mag) & $-0.21\pm0.04$ & $1.52\pm0.17$ & $0.000$*\\
        (j) & $\log\,(\rm sSFR/yr^{-1})$ & $0.33\pm0.06$ & $-8.88\pm0.11$ & $0.000$*\\
        (k) & $q$ & $0.13\pm0.07$ & $0.91\pm0.50$ & $0.594$\\
        (l) & $\log\, n^{\rm opt}$ & $2.24\pm5.14$ & $-2.83\pm4.19$ & $0.497$\\
        \hline
        \multicolumn{5}{c}{$\nabla = s(X-X_0)$}\\ [1.5ex]
        (m) & $\log\, (R_e^{\rm opt}/\rm kpc$) & $-0.36\pm0.14$ & $0.37\pm0.18$ & $0.010$*\\
        (n) & $\log\, (M_\star/ M_{\odot})$ & $-0.17\pm0.04$ & $10.00\pm0.28$ & $0.000$*\\
        (o) & UV$-$optical color (mag) & $-0.28\pm0.05$ & $1.25\pm0.13$ & $0.000$*\\
        (p) & $\log\,(\rm sSFR/yr^{-1})$ & $0.40\pm0.08$ & $-8.74\pm0.10$ & $0.000$*\\
        (q) & $q$ & $0.11\pm0.22$ & $-4.05\pm20.56$ & $0.825$\\
        (r) & $\log\, n^{\rm opt}$ & $0.67\pm0.33$ & $-15.88\pm190.34$ & $0.934$\\
    \enddata

    \tablecomments{
Col. (1): The label of the linear regression.
Col. (2): The galaxy property that serves as the independent variable $X$ in the regression.
Col. (3): The best-fit slope of the linear regression.
Col. (4): The typical value $X_0$ of the independent variable $X$ that makes the estimated variation (i.e. the estimated value of $\mathcal{R}$ or $\mathcal{N}$ at $X_0$) equals 1. For $\nabla$, $X_0$ refers to the typical value that makes $\nabla = 0$.
Col. (5): The probability of accepting the null hypothesis that the slope is zero (i.e. no statistical correlation). \\
$^*$We take $p$-value $\, < 0.05$ as the probability of rejecting the null hypothesis of no correlation.
    }
\end{deluxetable}

\begin{figure*}
\centering
\includegraphics[width=0.9\textwidth]{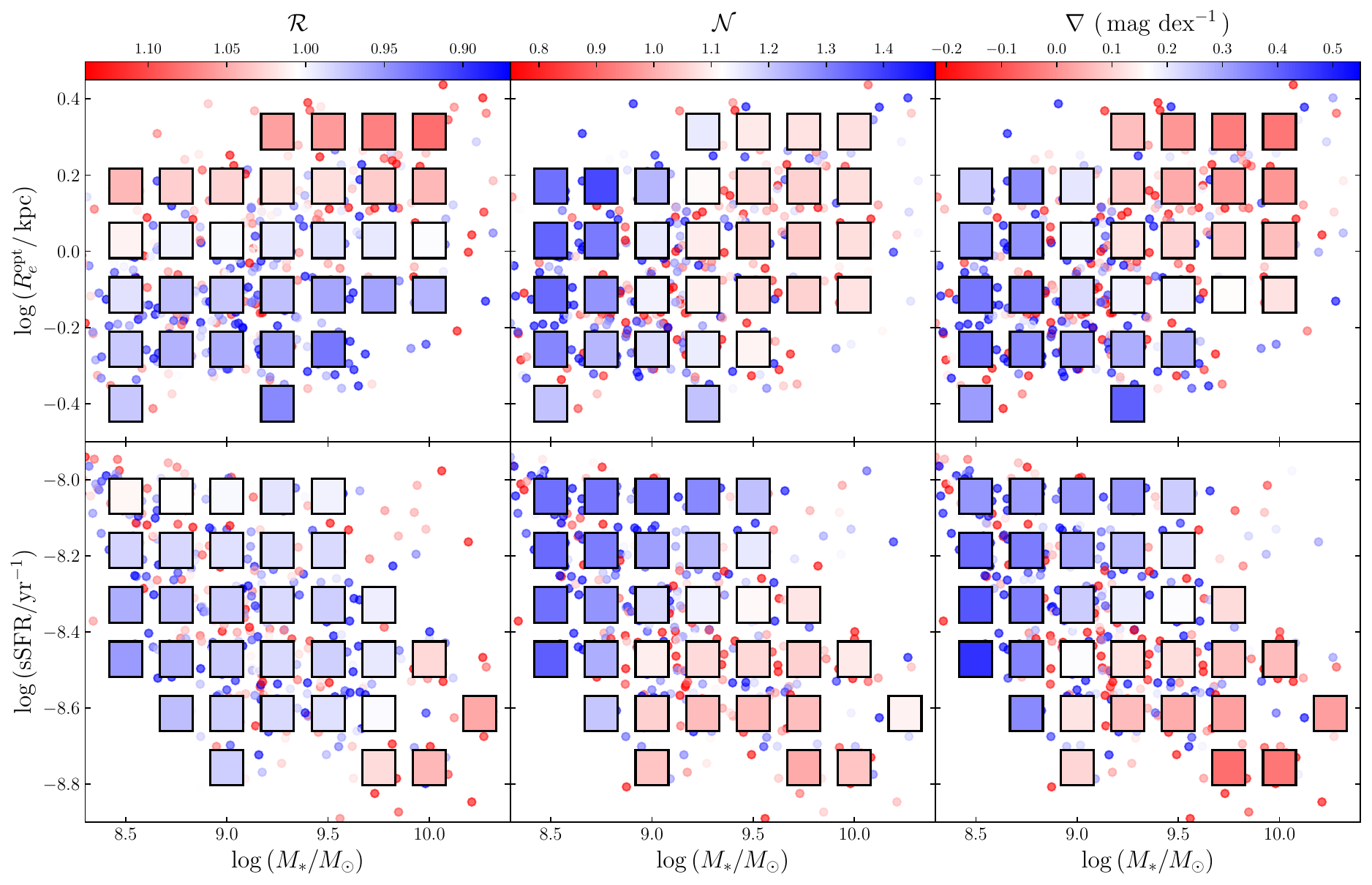}
\caption{The stellar mass versus size and stellar mass versus specific SFR (sSFR) distributions of our sample. From left to right, the columns quantify the color gradient using $\mathcal{R}$, $\mathcal{N}$, and $\nabla$, whose strength is indicated by the color bar on the top. The squares denote median values of at least five objects within a box of size $0.25$~dex in $\log M_\star$ and 0.11~dex in $\log R_e^{\rm opt}$ or 0.13 mag in rest-frame UV-to-optical color, after performing a locally weighted regression smoothing using the Python package \texttt{LOESS} \citep{CappellariLOESS2013MNRAS}. \label{fig:msmc}}
\end{figure*}

\subsection{Dependence on Galaxy Properties} \label{subsec:dependency}

Does the color gradient depend on galaxy properties? We investigate the trends between color gradient and galaxy size, integrated rest-frame UV$-$optical color, stellar mass, and other physical properties computed using \texttt{Bagpipes}\footnote{{\url{https://bagpipes.readthedocs.io/en/latest/}}.} \citep{Carnall2018}. We choose \texttt{Bagpipes} because it is a flexible SED-fitting code that has been widely used for high-redshift galaxies \citep[e.g.,][]{wangUNCOVERSurveyFirstLook2024,Champagne2025,Harvey2025}. \texttt{Bagpipes} can also incorporate a non-parametric star formation history, which has been adopted extensively in studies of high-redshift galaxies (e.g., \citealt{Tacchella2020}).

Following \citet{Leja2019}, we use a non-parametric star formation history with four time bins, where the first bin has a lookback time of 30 Myr and the oldest bin ends with 90\% of the age of the Universe at the source's redshift. We use a \cite{Bruzual2003} stellar population model with metallicity in the range $(0.01-2.0)\, Z_\sun$. Dust attenuation is treated using the law of \cite{Calzetti2000} with $E(B-V) = 0-2$ mag. We also include nebular emission with a fixed ionization parameter of $U = 10^{-3}$. The SED fits are performed on the combination of all NIRCam photometry (F115W, F150W, F200W, F277W, F356W, F410M, F444W) and ancillary photometry of bands not covered by JWST (HST/ACS F606W and F814W; Spitzer IRAC3 and IRAC4). We adopt photometric redshifts from CANDELS and a Galactic extinction of $E(B-V)=0.007$ mag with $R_V=3.1$ \citep{Schlafly2011}. When combining our new, more accurate NIRCam photometry with the extant panchromatic photometry of the EGS field \citep{stefanonCANDELSMultiwavelengthCatalogs2017}, we find it advantageous to boost the uncertainties of our measurements by an additional 10\%, a relative error determined by the typical systematic offset between our model fluxes and those from the publicly available catalog of \citet{Weaver2024}.

Figure~\ref{fig:dep} shows the dependence of the color gradient on galaxy properties. In each subpanel, we calculate the running median to accentuate the trend. Each bin includes at least 60 galaxies, and the error bars reflect the scatter of the trend, which is estimated from the standard deviation of the data points in each bin. 

We find a statistically significant correlation between $\mathcal{R}$ and galaxy size, characterized using the effective radius ($R_e^{\rm opt}$) measured at the optical band closest to rest-frame 5000~\AA, such that galaxies with larger size tend to have more negative color gradients (Figure~\ref{fig:dep}, top row). The trend is robust when considering effective radii measured in other bands (Appendix~\ref{app:size}). The dependence on stellar mass is less steep, but, with $p$-value = 0.027, formally still qualifies as statistically significant (Table~\ref{tab:fitresult}). The flattened slope can be explained by the relatively shallow relation between the size and stellar mass \citep{wardEvolutionSizeMassRelation2024} and luminosity \citep{onoCensusRestframeOptical2024, sunStructureMorphologyGalaxies2023} of star-forming galaxies. The trend with stellar mass naturally extends to the rest-frame UV$-$optical color, insofar as more massive galaxies are redder. The case for $\mathcal{N}$ is more ambiguous (Figure~\ref{fig:dep}, middle row), which is understandable because the S{\'e}rsic indices of faint, high-redshift galaxies are much more difficult to measure than their effective radii \citep{sunStructureMorphologyGalaxies2023}. Although $\mathcal{N}$ is formally uncorrelated with galaxy size, the dependence on stellar mass is statistically robust ($p$-value $< 0.001$), consistent with the tendency for high-mass systems to exhibit negative color gradients and low-mass systems to show the opposite. Again, the correlation with stellar mass extends to the rest-frame UV$-$optical color. Among the three estimators of the color gradient, the most pronounced trends emerge with $\nabla$ (Figure~\ref{fig:dep}, bottom row). 

Additional insight can be gleaned from considering the distribution of color gradients on the diagrams of stellar mass versus size and stellar mass versus specific SFR (sSFR), which are shown on the top and bottom rows of Figure~\ref{fig:msmc}. Here, it can be seen that at {\it fixed}\ stellar mass the color gradient depends on galaxy size: as the effective radius increases, the color gradient systematically shifts from positive to negative. These trends are most clearly seen when the color gradient is estimated using $\mathcal{N}$ or $\nabla$, but it is also apparent when $\mathcal{R}$ is used. Similarly, at a given stellar mass the color gradients vary systematically with sSFR: galaxies with more active star formation have blue centers and redder exteriors.

Lastly, we examine whether any connection can be found between color gradient and galaxy morphology. Using the S{\'e}rsic index in the rest-frame optical ($n^{\rm opt}$) as a crude guide to morphology, we find no statistically meaningful relation between $n^{\rm opt}$ and any of the three measures of color gradient (Table~\ref{tab:fitresult}). We do see a very tentative link between color gradient and galaxy axial ratio ($q$), but the statistics fail to prove its robustness.

In summary, both positive and negative color gradients exist in $z > 4$ galaxies, but the relative proportion of positive gradients becomes much more prevalent than at low redshifts. Notably, the sign of the color gradient depends systematically on galaxy properties. Whereas galaxies with negative color gradients are generally larger, more massive, and redder, positive color gradients preferentially occur in smaller, low-mass, bluer, possibly more disk-dominated or prolate systems. For any given stellar mass, more compact galaxies are more likely to have a positive color gradient.

\begin{figure}
\centering
\includegraphics[width=0.45\textwidth]{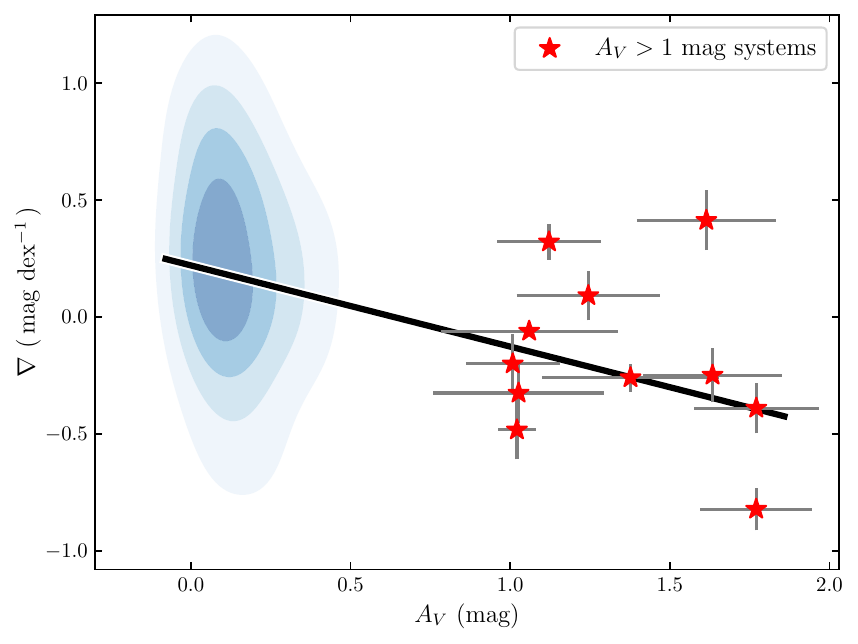}
\caption{The correlation between dust attenuation and $\nabla$ in our sample. Dusty systems ($A_V>1$ mag) are highlighted in red stars. The blue contours display the distribution of all sample galaxies, with the black solid line indicating the best-fit linear regression.\label{fig:dustrole}}
\end{figure}

\begin{figure}
\centering
\includegraphics[width=0.46\textwidth]{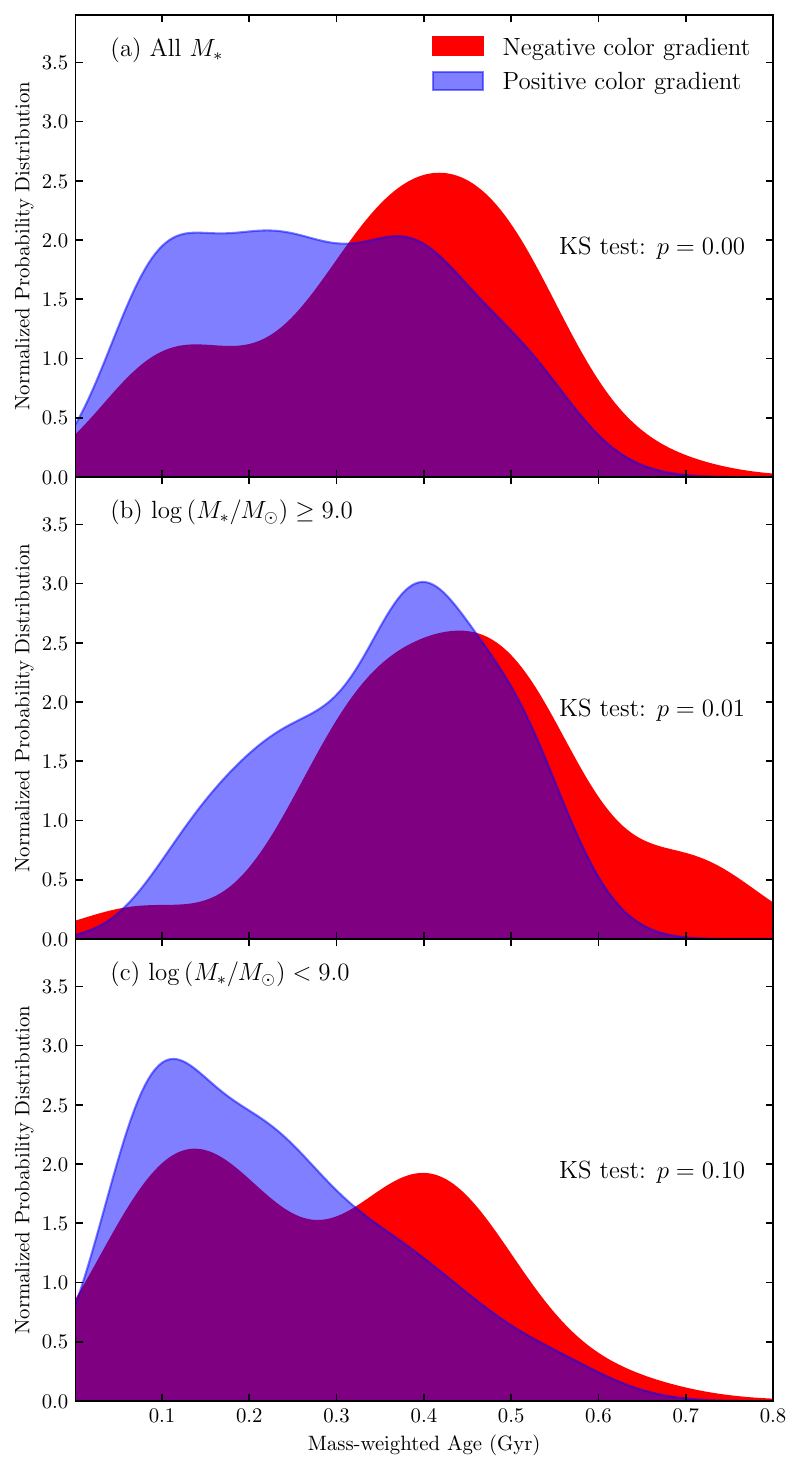}
\caption{The distribution of mass-weighted ages of galaxies with $A_V<0.4$ mag of (a) all stellar masses, (b) $M_\star \geq  10^9\,M_{\odot}$, and (c) $M_\star < 10^9\,M_{\odot}$ having negative (red) and positive (blue) color gradient. The $p$-value of the KS test is shown on the upper left corner of each panel.  \label{fig:age}}
\end{figure}

\section{Discussion} \label{sec:discuss}

\subsection{Origin of the Color Gradients} \label{subsec:excess}

Color gradients in galaxies principally arise from radial variations in stellar population, dust reddening, or both. To evaluate whether any particular factor dominates in shaping galaxy color gradients at high redshift, we examine the role of integrated dust attenuation ($A_V$) and the mass-weighted age derived from SED fitting. We separate the galaxies into three categories based on the magnitude of their color gradient: galaxies with $|\nabla|$ greater than its uncertainty are considered to have a robust negative or positive color gradient, while and those with $|\nabla|$ smaller than its uncertainty are regarded as inconclusive or having a flat color gradient. The inverse correlation between the color gradient and dust attenuation in Figure \ref{fig:dustrole} testifies to the importance of dust, but the effect is confined to a handful (11/441) of systems with $A_V>1$ mag. The vast majority (396/441) of the sample are clustered in a region of low attenuation ($A_V \lesssim 0.4$ mag.)

Isolating the low-attenuation sources reveals a bimodal distribution of mass-weighted age that seems to be correlated with color gradient (Figure \ref{fig:age}): galaxies with positive color gradients have an overall younger stellar population, while negative color gradients are statistically more closely associated with older galaxies. This trend agrees with the statistical correlation between color gradient and sSFR (Figure~\ref{fig:msmc}).  Intriguingly, a two-sample Kolmogorov-Smirnov (KS) test indicates that the age difference between the two populations is confined exclusively to massive galaxies ($M_* \geq 10^{9.0}\,M_\odot$; Figure~\ref{fig:age}b) and is hardly perceptible among less massive systems ($M_* < 10^{9.0}\,M_\odot$; Figure~\ref{fig:age}c).

Cosmological simulations of massive ($M_*\approx 10^{9}-10^{10}\,M_{\odot}$) galaxies at $z \approx 2$ produce ``blue nuggets'' with central star formation enhanced by violent disk instability, minor mergers, and counter-rotating streams \citep{zolotovCompactionQuenchingHighz2015}. This ``compaction'' phase of galaxy evolution has garnered support from earlier HST studies \citep[e.g.,][]{Barro2017, tacchella2018ApJ...859...56T, Suess2021, Ji2023}, and it has been reaffirmed by recent JWST/MIRI observations that spatially map dust-obscured star formation in $z\lesssim2.5$ galaxies \citep{magnelliCEERSMIRIDeciphers2023, Lyu2025}. Examples of ``outside-in'' growth also can be found much closer to home. For instance, \citet{tuttleBreakBRD2020} identified a group of 118 $z<0.05$ ``BreakBRDs'' (break bulges in red disks) galaxies that feature truncated disk star formation and a blue star-forming core, whose central concentration of gas and subsequent star formation may be traced plausibly to the low angular momentum of their circumgalactic medium \citep{TonnesenCGM2023,StarkBreakBRD2024}. Analyzing the half-light radii measured from HST F275W and F160W images of $0.05 < z < 0.3$ galaxies selected from the GOODS-North field, \citet{chengUVNIRSize2020} concluded that while high-mass galaxies conform with the expectations of the conventional inside-out growth mode, some galaxies with $M_* < 10^{8}\,M_{\odot}$ display small-scale UV structures consistent with ongoing central star formation.
 
The high incidence of positive color gradients revealed in our study is qualitatively consistent with the prevalence of compact UV morphology reported in $z \gtrsim 5$ galaxies \citep[e.g.,][]{ morishitaEnhancedSubkpcScale2024,harikane2025}. The compactness of these UV-bright systems suggests that star formation proceeds through a violent central starburst, possibly triggered by angular momentum dissipation due to mergers or in situ instabilities \citep{dekelMassThresholdGalactic2020}. Moreover, we find that positive color gradients preferentially occur in low-mass ($M_{\star}\lesssim 10^9\, M_{\odot}$) galaxies. A similar pattern is reported for the $4\leq z\leq 10$ galaxies studied by \citet{tripodiSpatiallyResolvedEmission2024}, whose emission-line equivalent widths are most intense in the centers of low-mass systems. Their stacking analysis also tentatively suggests that the metallicity gradient in these galaxies is inverted. Low-mass galaxies may be particularly prone to centrally concentrated star formation because their gas depletion time is shorter than the replenishment time, and common merger-driven spin flips \citep{Tacchella2016, dekelMassThresholdGalactic2020} facilitate inward mass transport. Moreover, the apparent connection between positive color gradients and smaller axis ratios, although statistically only marginal (Table~\ref{tab:fitresult}), might arise naturally if low-mass galaxies are initially prolate, a notion advanced by cosmological simulations \citep{LapinerBN2023} and supported by observations \citep{vanderwelProlate2014,zhang3DIntrinsicShapes2019, pandyaGalaxiesGoingBananas2023}.

\begin{figure}
\centering
\includegraphics[width=0.98\linewidth]{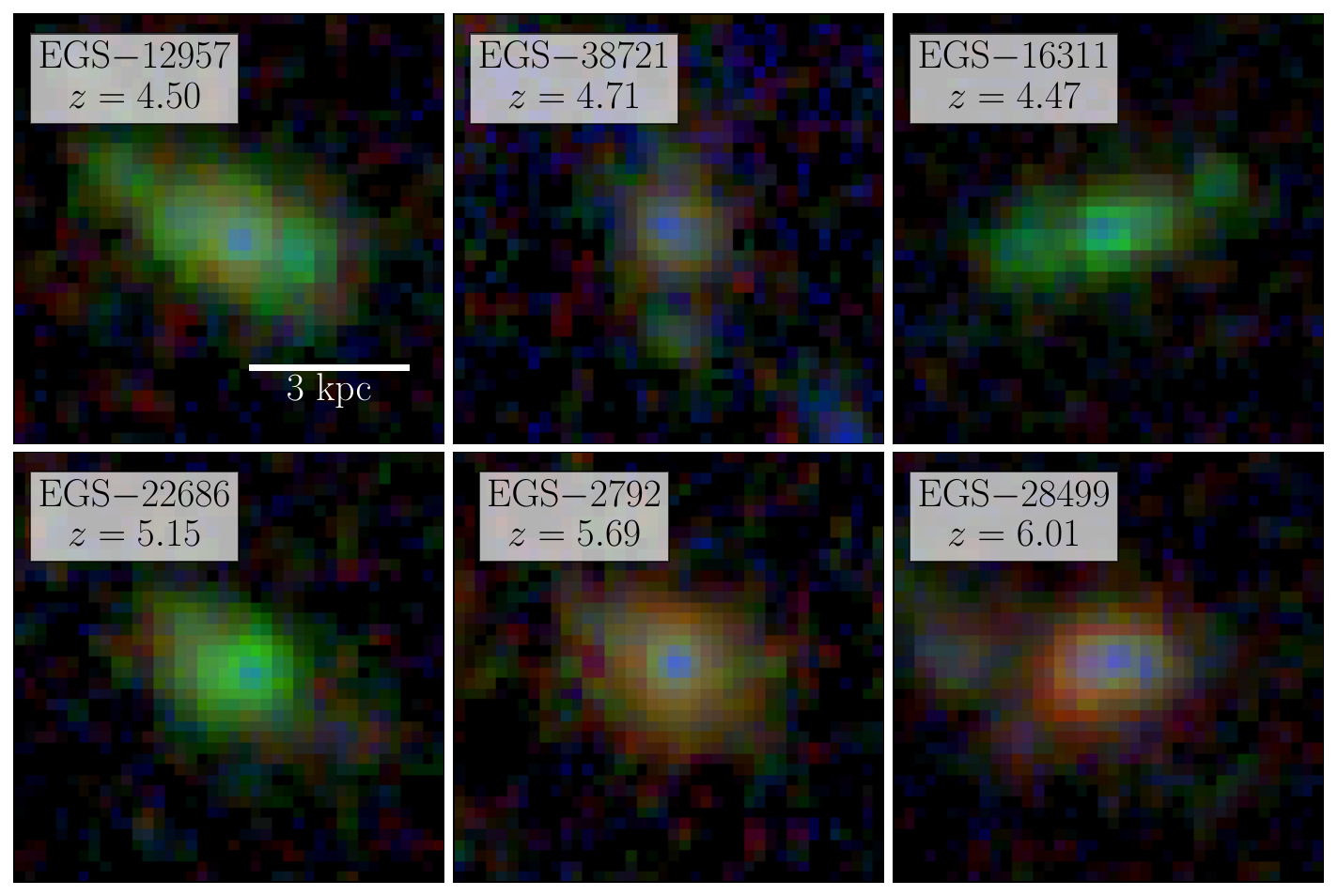}
\caption{Pseudo-color (red = F444W, green = F277W, blue = F150W) images of galaxies with positive color gradients that show evidence of hosting a central point-like source.
\label{fig:pls}}
\end{figure}

\begin{figure*}
\centering
\includegraphics[width=0.98\linewidth]{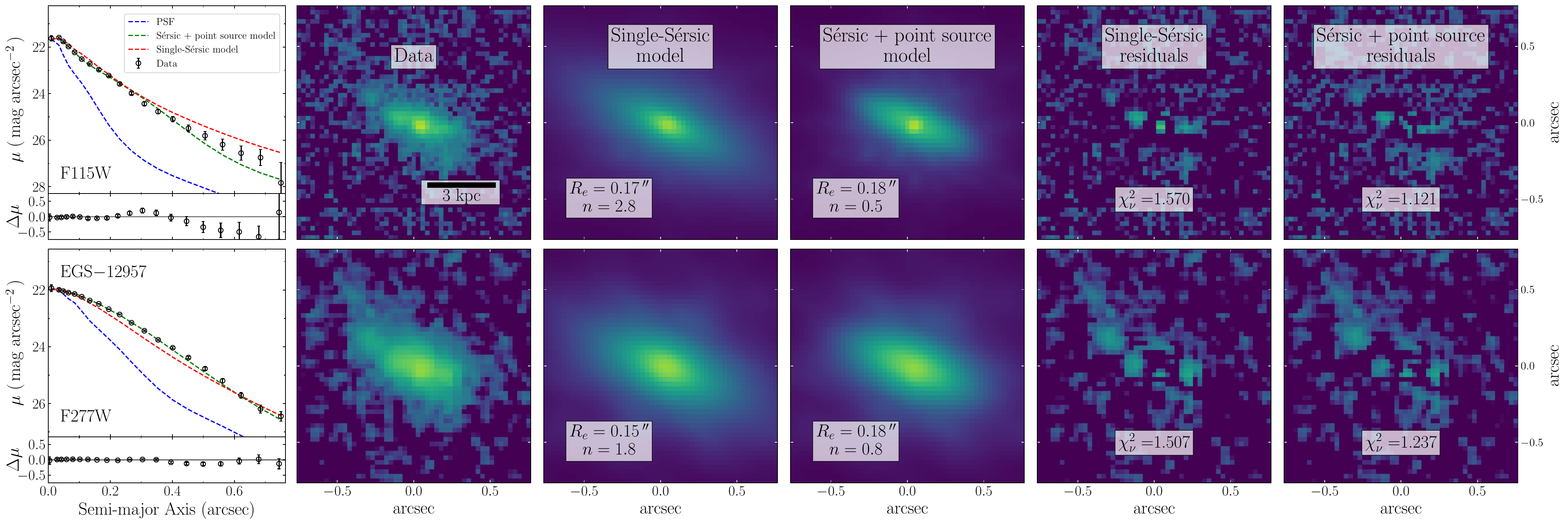}
\caption{Influence of a ``blue nucleus'' on the measurements, for F150W (top) and F444W (bottom). In the surface brightness profile, the black points with error bars are measured by aperture photometry, the red dashed line shows the single-S\'ersic model, the green dashed line refers to the S\'ersic+point source model, and the blue dashed line represents the PSF, scaled to match the model profile. The lower panel of the left column shows the residuals between the S{\'e}rsic+point source model and the data. The images show, from left to right, the original image, the single-S\'ersic model, the S\'ersic+point source model, and the respective residuals for the two models. It is clear that including the nucleus component improves the quality of the fit, and that the additional component is necessary to measure accurately the structural parameters of the host galaxy. \label{fig:galfitpls}}
\end{figure*}

\subsection{The Role of Active Galactic Nuclei} \label{agn}

Active galactic nuclei (AGNs) complicate the interpretation of our results. To the extent that a nuclear point source, if unobscured, boosts the optical-UV flux of the central region and artificially alters the structural parameters of the morphological analysis, an AGN can be confused with compact star formation. This is especially worrisome in light of the unexpectedly high abundance of high-redshift AGNs discovered in the JWST era \citep[e.g.][]{KokorevLRD2024ApJ, Labbe2025}. Careful inspection of the results of our two-dimensional image analysis reveals that 23 galaxies (5\% of \nsample~galaxies) are better fit by adding an additional point-source component to the underlying S\'ersic model for the galaxy. These are flagged as potential AGN host galaxies\footnote{The true physical nature of the central point source remains ambiguous, however, in the absence of reliable spectroscopic information.} in Table~\ref{tab:format}. Most of them (16/23; 69\%) exhibit a positive color gradient, the rest negative. Figure~\ref{fig:pls} showcases some galaxies whose central colors are clearly influenced by a prominent blue core, which is likely to be an active nucleus. One of the examples, EGS$-$28499, refers to the spectroscopically identified broad-line AGN CEERS\,00397 \citep{harikaneJWSTNIRSpecFirst2023}. We identify three additional AGN candidates by cross-matching our sample with those studied by other methods, including spectroscopic analysis \citep{harikaneJWSTNIRSpecFirst2023}, SED fitting  \citep{Park+2010+irselectedagn, Onouemasafusa+2023+agn}, and X-ray observations \citep{Kocevski+2023+AGN}. The presence of an AGN core can distort substantially the inferred color gradient of the underlying host galaxy. An example is given in Figure~\ref{fig:galfitpls}, which shows that in the case of EGS$-$12957 removing the nuclear point source drastically affects the S\'ersic index of the galaxy, especially in the F115W band, such that its color gradient proxy switches from an original value of $\mathcal{N} = 1.6$ to a nucleus-removed value of $\mathcal{N} = 0.6$, flipping the color gradient from positive to negative. It is important to note that excluding the AGN candidates (flagged in Table~\ref{tab:format}) from our analysis does not affect the main conclusions of this study.

\subsection{Comparison with Previous Studies} \label{subsec:comparison}

Galaxies at low redshifts are generally significantly larger in the rest-frame UV than in the optical because of the preponderance of negative color gradients \cite[e.g.,][]{vanderwel3DHSTCANDELSEvolution2014, shibuyaMorphologies1900002015}. In their investigation of galaxies across the main sequence at $z\approx 0.7-1.5$, \citet{nelsonWhereStarsForm2016} argued that the increasing size discrepancy between the H$\alpha$ and stellar continuum morphology constitutes evidence of inside-out growth. This trend becomes less pronounced at $z \approx 1-3$ \citep{suessHalfmassRadii0002019}, and by $z \gtrsim 4$ we find that an increasingly large fraction of galaxies sport positive instead of negative color gradients. Similar results were obtained by \citet{morishitaEnhancedSubkpcScale2024} in their study of the UV and optical sizes of 149 $z \geq 4$ galaxies, by \citet{JiaSizeGrowth2024} with the comparison of rest-frame 0.45 $\mu$m and 1.0 $\mu$m sizes for $z\gtrsim4$ galaxies in JADES, as well as by \citet{treuEarlyResultsGLASSJWST2023} for their sample of 19 Lyman-break galaxies at $z>7$. Interestingly, the sign and magnitude of the gradient depend on galaxy mass (and thus, indirectly, on galaxy size), such that more massive galaxies exhibit steeper negative color gradients than low-mass systems. \citet{suessHalfmassRadii0002019} conclude that the strength of the negative color gradient is stronger for galaxies of larger size, higher stellar mass, and redder color, precisely the trends found here for galaxies at $z > 4$. In contrast to the inside-out growth for massive systems, galaxies of intermediate stellar mass ($M_\star\approx 10^{9}\, M_\odot$) tend to have relatively flat color profiles, an effect attributed to coherent star formation across all radii in this mass regime \citep{vandokkum2013ApJ...771L..35V, tacchella2018ApJ...859...56T, nelson2021MNRAS.508..219N, Hasheminia2024}. And by the time we reach even lower masses ($M_\star\lesssim 10^{9}\, M_\odot$), positive color gradients outnumber negative ones.

\section{Summary} \label{sec:summary}

We conduct a comprehensive study of the color gradients of \nsample~high-redshift ($4<z<8$) galaxies using the NIRCam images from the JWST CEERS program. The structural parameters of the galaxies are obtained by simultaneously fitting the seven-band images using the two-dimensional analysis code \texttt{GalfitM}. We quantify the color gradient using three empirical proxies ($\mathcal{R},\mathcal{N}, \nabla$) that can be derived robustly from the parametric fits, whose uncertainties can be inferred from the extensive mock simulations performed by \cite{sunStructureMorphologyGalaxies2023}. The main results are summarized as follows:

\begin{enumerate}

\item Both positive and negative color gradients are present in $z>4$ galaxies, with the frequency of positive gradients ($\sim 50\%-60\%$) markedly higher than found at lower redshifts ($\sim 10\%-20\%$). Within the limitations of the current sample size, we discern no evidence of redshift evolution in color gradient within the range $4<z<8$.

\item As the majority of the galaxies in our sample are not strongly attenuated by dust, their color gradients principally reflect radial variations in stellar population. While emission from a point-like active nucleus can present a significant source of contamination and confusion in the interpretation of the color gradients, we argue that our main conclusions are robust against AGN contamination.

\item The sign and magnitude of the color gradient depend systematically on the global properties of the galaxy, such that galaxies of higher stellar mass, larger size, and redder integrated color possess increasingly steeper negative color gradients indicative of inside-out growth. Galaxies of lower stellar mass, here defined by $M_\star\lesssim 10^{9}\, M_\odot$, more compact size, and bluer colors have preferentially positive color gradients, consistent with centrally concentrated star formation.

\end{enumerate}

Future work can be improved by enlarging the sample using other JWST surveys, such as COSMOS-Web \citep{caseyCOSMOSWebOverviewJWST2023}, JADES \citep{eisensteinOverviewJWSTAdvanced2024}, and NGDEEP \citep{bagleyNextGenerationDeep2023}. Although we have ruled out the possibility that nuclear activity is the main culprit of the excess positive color gradients in our sample, we are still unsure to what extent AGNs influence the internal properties of galaxies at high redshifts. Adding spectral information, which will become available for increasingly large JWST samples, will be highly informative.

\begin{acknowledgments}
We are grateful to the referee for helpful comments and suggestions. This research was supported by the National Science Foundation of China (12233001) and the China Manned Space Program (CMS-CSST-2025-A09). We thank Chang-Hao Chen, Fangzhou Jiang, Limin Lai, Ruancun Li, Yang Li, Zhao-Yu Li, Chao Ma, Jinyi Shangguan, and Ming-Yang Zhuang for discussions and technical advice.
\end{acknowledgments}

\vspace{5mm}
\facilities{JWST (NIRCam)}
\software{{\tt Astropy} \citep{theastropycollaborationAstropyCommunityPython2013},
{\tt Bagpipes} \citep{Carnall2018},
{\tt Galfit} \citep{pengDetailedStructuralDecomposition2002,pengDETAILEDDECOMPOSITIONGALAXY2010},
{\tt GalfitM} \citep{hausslerMegaMorphMultiwavelengthMeasurement2013,vikaMegaMorphMultiwavelengthMeasurement2013},
{\tt Photutils} \citep{bradleyAstropyPhotutils122024},
{\tt Matplotlib} \citep{hunterMatplotlib2DGraphics2007},
{\tt NumPy} \citep{harrisArrayProgrammingNumPy2020},
{\tt SciPy} \citep{virtanenSciPyFundamentalAlgorithms2020}}

\appendix

\section{Color Gradient and SED-fitting Analysis} \label{app:table}

Table~\ref{tab:format} presents the catalog of parameters derived from the color gradient and SED-fitting analysis.

\renewcommand*{\thetable}{A1}
\begin{longtable}{cccc}
\caption{Catalog of Structural and Physical Parameters}\label{tab:format}\\
\hline \hline \\[-2ex]
   \multicolumn{1}{c}{\textbf{Column}} &
   \multicolumn{1}{c}{\textbf{Format}} &
   \multicolumn{1}{c}{\textbf{Name}} &
   \multicolumn{1}{c}{\textbf{Description}} \\[0.5ex] \hline
   \\[-1.8ex]
\endfirsthead
\multicolumn{4}{c}{{\tablename} \thetable{} -- Continued} \\[0.5ex]
  \hline \hline \\[-2ex]
  \multicolumn{1}{c}{\textbf{Column}} &
  \multicolumn{1}{c}{\textbf{Format}} &
  \multicolumn{1}{c}{\textbf{Name}} &
  \multicolumn{1}{c}{\textbf{Description}} \\[0.5ex] \hline
  \\[-1.8ex]
\endhead
  \hline
  \multicolumn{4}{l}{{Continued on Next Page\ldots}} \\
\endfoot
  \\[-1.8ex] \hline \hline
\endlastfoot
1\dotfill  & STRING & Name & Object name from Stefanon et al. (2017)\\
2\dotfill   & DOUBLE & RAdeg & Right ascension in decimal degrees (J2000.0) \\
3\dotfill   & DOUBLE & DECdeg & Declination in decimal degrees (J2000.0) \\
4\dotfill   & DOUBLE & z & Redshift \\
5\dotfill   & DOUBLE & logMs & Stellar mass from SED fitting\\
6\dotfill   & DOUBLE & e\_logMs & Mean error on $\log\,M_\star$ \\
7\dotfill   & DOUBLE & logsSFR & Specific Star formation rate from SED fitting \\
8\dotfill   & DOUBLE & e\_logsSFR & Mean error on $\log\,\rm{sSFR}$ \\
9\dotfill   & DOUBLE & age & Mass-weighted age from SED fitting \\
10\dotfill   & DOUBLE & e\_age & Mean error on mass-weighted age \\
11\dotfill   & DOUBLE & Av & Dust attenuation from SED fitting \\
12\dotfill   & DOUBLE & e\_Av & Mean error on dust attenuation \\
13\dotfill & DOUBLE & q & Axis ratio $q$\\
14\dotfill & DOUBLE & e\_q & Mean error on $q$\\
15\dotfill & DOUBLE & PA  & Position angle $\Theta$\\
16\dotfill & DOUBLE & e\_PA  & Mean error on $\Theta$\\
17\dotfill & STRING & RestFrameUVBand  & Rest-frame UV band\\
18\dotfill & DOUBLE & ReUV & Half-light radius $R_e$ in the rest-frame UV band \\
19\dotfill & DOUBLE & e\_ReUV & Mean error on ReUV\\
20\dotfill & DOUBLE & nUV & S{\'e}rsic index $n$ in the rest-frame UV band\\
21\dotfill & DOUBLE & e\_nUV & Mean error on nUV\\
22\dotfill & STRING & RestFrameOpticalBand  & Rest-frame Optical band\\
23\dotfill & DOUBLE & ReOpt  & Half-light radius $R_e$ in the rest-frame optical band \\
24\dotfill & DOUBLE & e\_ReOpt & Mean error on ReOpt\\
25\dotfill & DOUBLE & nOpt & S{\'e}rsic index $n$ in the rest-frame optical band\\
26\dotfill & DOUBLE & e\_nOpt & Mean error on nOpt\\
27\dotfill & DOUBLE & R & Size ratio of rest-frame UV to optical ($\mathcal{R}$)\\
28\dotfill & DOUBLE & e\_R & Mean error on $\mathcal{R}$\\
29\dotfill & DOUBLE & N & S\'ersic Index ratio of rest-frame UV to optical ($\mathcal{N}$)\\
30\dotfill & DOUBLE & e\_N & Mean error on $\mathcal{N}$\\
31\dotfill & DOUBLE & Nabla & Two-point slope of UV-optical color profile (gradient $\nabla$)\\
32\dotfill & DOUBLE & e\_Nabla & Mean error on $\nabla$\\
33\dotfill & DOUBLE & ChisqGalfitM    & Reduced chi-square from \texttt{GalfitM}\\
34\dotfill & LONG & FlagBLD & Visual degree of blending\\
35\dotfill & LONG & FlagHOST & Possible AGN host galaxy flag\\
36\dotfill & DOUBLE & F115Wfnu & Flux in F115W\\
37\dotfill & DOUBLE & e\_F115Wfnu & Mean error on flux in F115W\\
38\dotfill & DOUBLE & F150Wfnu & Flux in F150W\\
39\dotfill & DOUBLE & e\_F150Wfnu & Mean error on flux in F150W\\
40\dotfill & DOUBLE & F200Wfnu & Flux in F200W\\
41\dotfill & DOUBLE & e\_F200Wfnu & Mean error on flux in F200W\\
42\dotfill & DOUBLE & F277Wfnu & Flux in F277W\\
43\dotfill & DOUBLE & e\_F277Wfnu & Mean error on flux in F277W\\
44\dotfill & DOUBLE & F356Wfnu & Flux in F356W\\
45\dotfill & DOUBLE & e\_F356Wfnu & Mean error on flux in F356W\\
46\dotfill & DOUBLE & F410Mfnu & Flux in F410M\\
47\dotfill & DOUBLE & e\_F410Mfnu & Mean error on flux in F410M\\
48\dotfill & DOUBLE & F444Wfnu & Flux in F444W\\
49\dotfill & DOUBLE & e\_F444Wfnu & Mean error on flux in F444W\\
\end{longtable}\scriptsize
\normalsize

\section{Mock Tests of the Color Gradient Proxies} \label{app:mock}

We summarize the mock analysis described in Section~\ref{subsec:uncertainty}, by illustrating some typical examples. For simplicity, we only illustrate using F150W as the blue band and F444W as the red band. These scenarios span a range of parameter space that covers our sample well. For each of the three color gradient proxies (Equations~\ref{eq:nvar}--\ref{eq:nabla}), we study the possible bias, defined as the difference between the input and the output parameter values, for three representative input values, as a function of galaxy size (represented by $R_e^{\rm F444W}$) and source brightness (represented by F150W magnitude). We consider four bins of galaxy size ($R_e^{\rm F444W} < 0\farcs032$, $0\farcs032 - 0\farcs060$, $0\farcs060 -0\farcs112$, $0\farcs112 - 0\farcs161$), six values of the S\'ersic index ($n = 0.71, 1.09, 1.66, 2.53, 3.86, 5.90$), and 10 bins in source brightness spanning $\rm F150W = 20-29$~mag.

\begin{figure}
\figurenum{B1}
\centering
\includegraphics[width=0.88\linewidth]{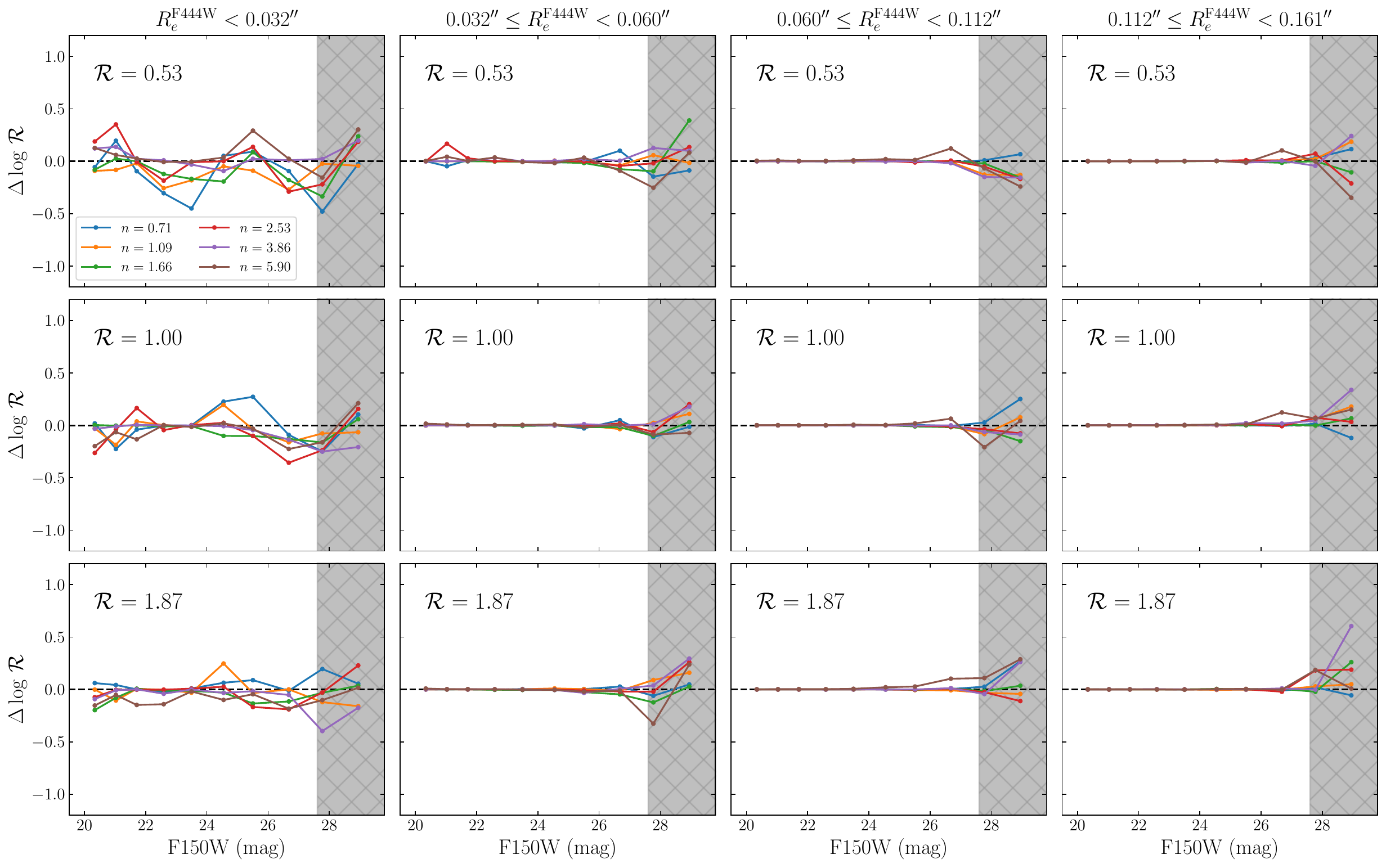}
\caption{The bias of $\mathcal{R}$ as a function of F150W magnitude, for $\mathcal{R} = 0.53$ (top), 1.00 (middle), and 1.87 (bottom), and, for columns from left to right, different values in input $R_e^{\rm F444W}$. The colors indicate the input values of the S{\'e}rsic index $n$ at F444W. The grey-shaded region marks the magnitude range below $5\,\sigma$ detection at HST/WFC3 F160W. \label{fig:mockR}
}
\end{figure}

\begin{figure}
\centering
\figurenum{B2}
\includegraphics[width=0.88\linewidth]{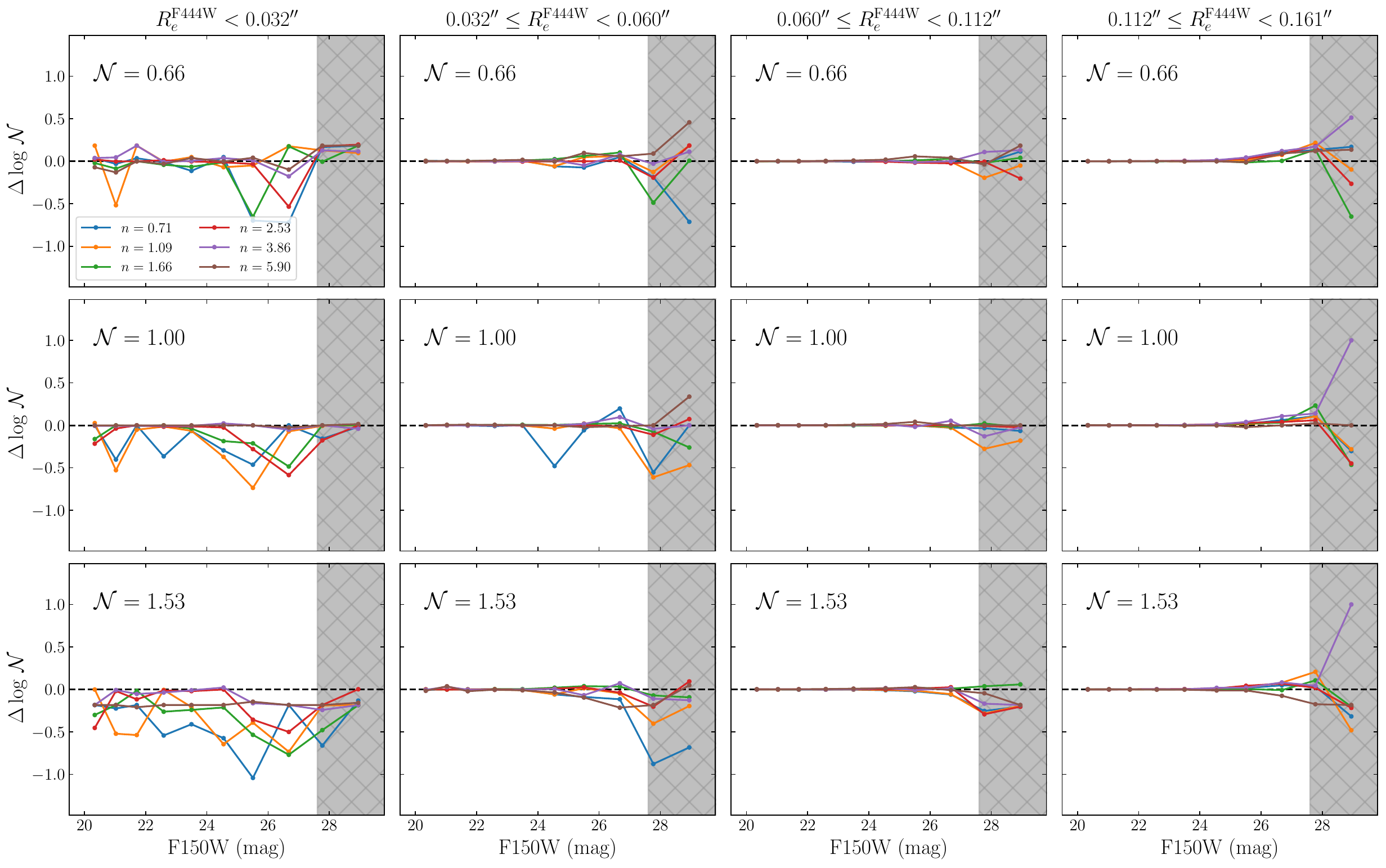}
\caption{As in Figure~\ref{fig:mockR}, but for the bias of $\mathcal{N}$, for $\mathcal{N} = 0.66$ (top), 1.00 (middle), and 1.53 (bottom).
\label{fig:mockN}
}
\end{figure}

\begin{figure}
\figurenum{B3}
\centering
\includegraphics[width=0.88\linewidth]{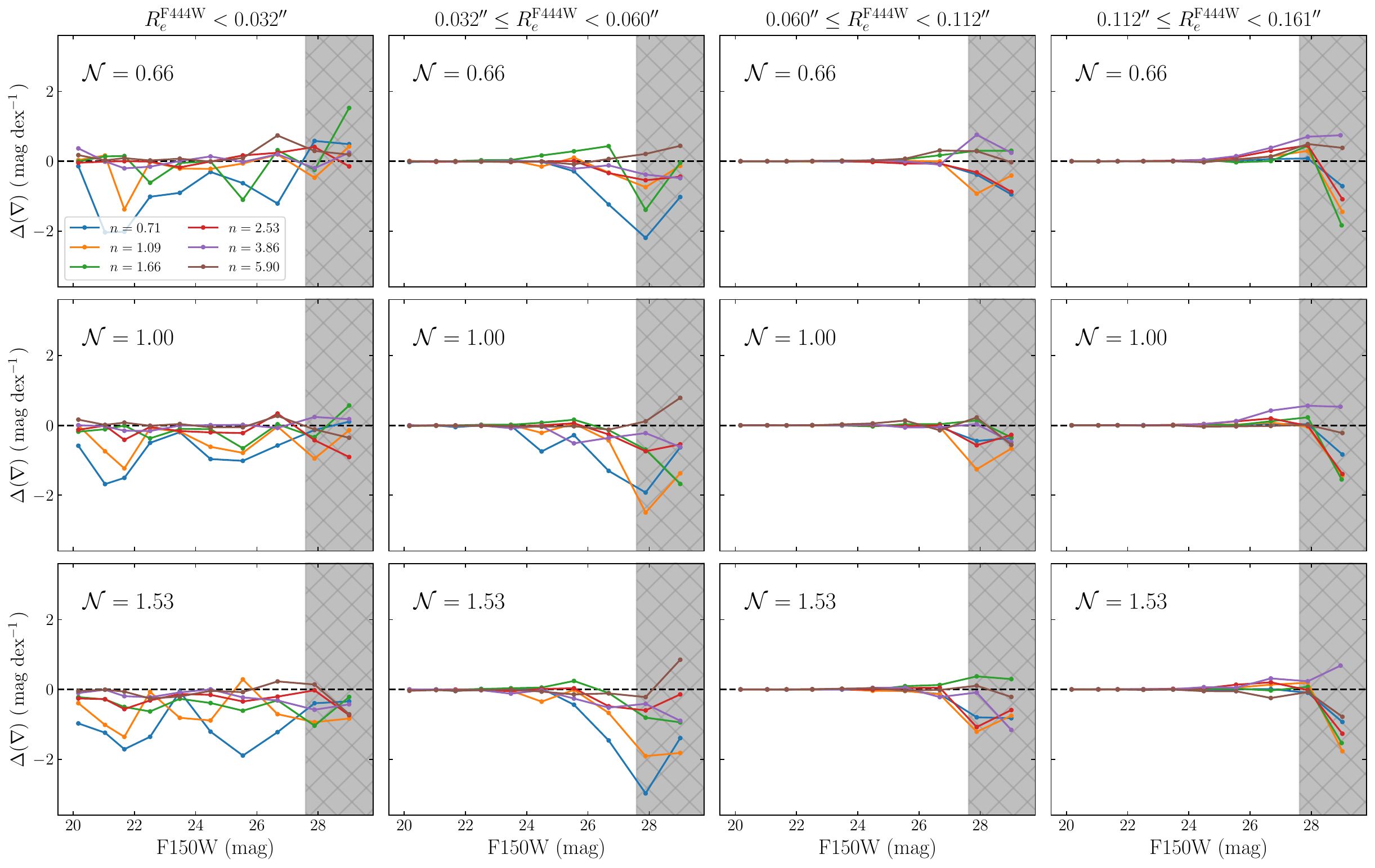}
\caption{As in Figure~\ref{fig:mockR}, but for the bias of $\nabla$, for $\mathcal{N} = 0.66$ (top), 1.00 (middle), and 1.53 (bottom). \label{fig:mockgrad}
}
\end{figure}

For the proxy that describes the wavelength variation of the effective radius, we choose $\mathcal{R}=0.53$, 1.00, and 1.87 as representative of the observed parameter space. Figure~\ref{fig:mockR} shows that $\Delta \log \mathcal{R}$ can be measured up to $\sim 0.21$~dex for all but the first size bin ($R_e^{\rm F444W} < 0\farcs032$) and for $\rm F150W \gtrsim 28$~mag. As for the proxy that describes the wavelength variation of the S\'ersic index, we choose $\mathcal{N}=0.66$, 1.00, and 1.53 as representative of the observed parameter space. Similarly, Figure~\ref{fig:mockN} shows that $\mathcal{N}$ is accurate to $\Delta \log \mathcal{N} \lesssim 0.29$ dex for all but $R_e^{\rm F444W} \lesssim 0\farcs060$ and $\rm F150W \gtrsim 28$~mag. The overall assessment for $\mathcal{N}$ also applies to $\nabla$ (Figure~\ref{fig:mockgrad}). We conclude that reliable measurements can be obtained for all three proxies of the color gradient so long as $R_e^{\rm F444W} > 0\farcs06$ and ${\rm F150W} < 28$~mag.
 
\section{Robustness of the Dependence of $\mathcal{R}$ on Galaxy Size} \label{app:size}

\begin{figure*}
\figurenum{C1}
\centering
\includegraphics[width=0.96\textwidth]{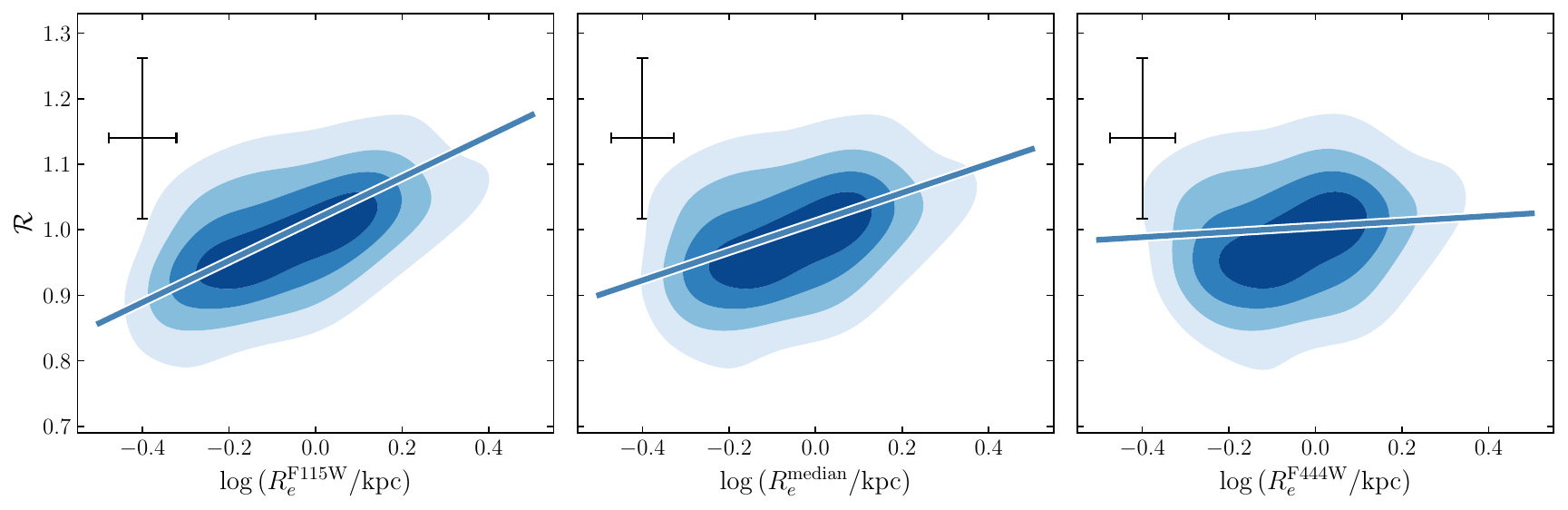}
\caption{Dependence of the variation in effective radius $\mathcal{R}$ on galaxy effective radius $R_e$ measured in F115W (left), the median of different bands (middle), and F444W (right). The solid line gives the best-fit linear regression. The error bars indicate the median uncertainty of the data points. The blue-filled contour shows the density distribution of the data points, where the darker color signifies higher density. \label{fig:bands}}
\end{figure*}

Section~\ref{subsec:dependency} (Figure~\ref{fig:dep}, top row) describes the inverse correlation between $\mathcal{R}$ and galaxy size, which was represented by $R_e^{\rm opt}$, the effective radius measured at the optical band closest to rest-frame 5000~\AA. Figure~\ref{fig:bands} verifies that the inverse correlation remains significant for different choices of effective radius, although the slope becomes shallower toward effective radii measured at longer wavelength.

~\\
~\\
~\\
~\\
~\\
~\\
~\\

\end{CJK*}

\begin{thebibliography}{}
\expandafter\ifx\csname natexlab\endcsname\relax\def\natexlab#1{#1}\fi
\providecommand{\url}[1]{\href{#1}{#1}}
\providecommand{\dodoi}[1]{doi:~\href{http://doi.org/#1}{\nolinkurl{#1}}}
\providecommand{\doeprint}[1]{\href{http://ascl.net/#1}{\nolinkurl{http://ascl.net/#1}}}
\providecommand{\doarXiv}[1]{\href{https://arxiv.org/abs/#1}{\nolinkurl{https://arxiv.org/abs/#1}}}
    

\bibitem[{{Abdurro'uf} \& Akiyama(2018)}]{abdurroufEvolutionSpatiallyResolved2018}
{Abdurro'uf}, \& Akiyama, M. 2018, \mnras, 479, 5083

\bibitem[{Abdurro'uf {et~al.}(2023)Abdurro'uf, Coe, Jung, Ferguson, Brammer, Iyer, Bradley, Dayal, Windhorst, Zitrin, Meena, Oguri, Diego, Kokorev, Dimauro, Adamo, Conselice, Welch, Vanzella, Hsiao, Xu, Roy, \& Mulcahey}]{abdurroufSpatiallyResolvedStellar2023} Abdurro'uf, Coe, D., Jung, I., {et~al.} 2023, \apj, 945, 117

\bibitem[{Bagley {et~al.}(2023)Bagley, Finkelstein, Koekemoer, Ferguson, Haro, Dickinson, Kartaltepe, Papovich, {P{\'e}rez-Gonz{\'a}lez}, Pirzkal, Somerville, Willmer, Yang, Yung, Fontana, Grazian, Grogin, Hirschmann, Kewley, Kirkpatrick, Kocevski, Lotz, Medrano, Morales, Pentericci, Ravindranath, Trump, Wilkins, Calabr{\`o}, Cooper, Costantin, {de la Vega}, Hutchison, Lucas, McGrath, Wang, \& Wuyts}]{bagleyCEERSEpochNIRCam2023} Bagley, M.~B., Finkelstein, S.~L., Koekemoer, A.~M., {et~al.} 2023, \apjl, 946, L12

\bibitem[Bagley et al.(2024)]{bagleyNextGenerationDeep2023} Bagley, M.~B., Pirzkal, N., Finkelstein, S.~L., et al.\ 2024, \apjl, 965, L6

\bibitem[Baker et al.(2024)]{bakerInsideoutGrowthEarly2024} Baker, W.~M., Tacchella, S., Johnson, B.~D., et al.\ 2024, Nature Astronomy

\bibitem[Barro et al.(2017)]{Barro2017} Barro, G., Faber, S.~M., Koo, D.~C., et al.\ 2017, \apj, 840, 1, 47

\bibitem[{Bradley {et~al.}(2024)Bradley, Sip{\H o}cz, Robitaille, Tollerud, Vin{\'i}cius, Deil, Barbary, Wilson, Busko, Donath, G{\"u}nther, Cara, Lim, Me{\ss}linger, Burnett, Conseil, Droettboom, Bostroem, Bray, Bratholm, Jamieson, Ginsburg, Barentsen, Craig, Pascual, Rathi, Perrin, Morris, \& Perren}]{bradleyAstropyPhotutils122024} Bradley, L., Sip{\H o}cz, B., Robitaille, T., {et~al.} 2024, Astropy/Photutils: 1.12.0, Zenodo

\bibitem[Bruzual \& Charlot(2003)]{Bruzual2003} Bruzual, G. \& Charlot, S.\ 2003, \mnras, 344, 4, 1000

\bibitem[Calzetti et al.(2000)]{Calzetti2000} Calzetti, D., Armus, L., Bohlin, R.~C., et al.\ 2000, \apj, 533, 2, 682 

\bibitem[Cappellari et al.(2013)]{CappellariLOESS2013MNRAS} Cappellari, M., McDermid, R.~M., Alatalo, K., et al.\ 2013, \mnras, 432, 1862

\bibitem[Carnall et al.(2018)]{Carnall2018} Carnall, A.~C., McLure, R.~J., Dunlop, J.~S., et al.\ 2018, \mnras, 480, 4, 4379

\bibitem[Carton et al.(2018)]{Carton+2018} Carton, D., Brinchmann, J., Contini, T., et al.\ 2018, \mnras, 478, 4293

\bibitem[Casey et al.(2023)]{caseyCOSMOSWebOverviewJWST2023} Casey, C.~M., Kartaltepe, J.~S., Drakos, N.~E., et al.\ 2023, \apj, 954, 31

\bibitem[Champagne et al.(2025)]{Champagne2025} Champagne, J.~B., Wang, F., Yang, J., et al.\ 2025, \apj, 981, 2, 114

\bibitem[{Cheng {et~al.}(2020)Cheng, Xu, Xie, Pan, Du, {Orellana-Gonz{\'a}lez}, Shao, Wu, Leiton, Huang, Dai, Assmann, \& Araneda}]{chengUVNIRSize2020} Cheng, C., Xu, C.~K., Xie, L., {et~al.} 2020, \aap, 633, A105

\bibitem[Ciotti(1991)]{Ciotti1991} Ciotti, L. 1991, \aap, 249, 99

\bibitem[{Dekel {et~al.}(2020)Dekel, Ginzburg, Jiang, Freundlich, Lapiner, Ceverino, \& Primack}]{dekelMassThresholdGalactic2020} Dekel, A., Ginzburg, O., Jiang, F., {et~al.} 2020, \mnras, 493, 4126

\bibitem[{Eisenstein {et~al.}(2025)Eisenstein, Willott, Alberts, Arribas, Bonaventura, Bunker, Cameron, Carniani, Charlot, {Curtis-Lake}, D'Eugenio, Endsley, Ferruit, Giardino, Hainline, Hausen, Jakobsen, Johnson, Maiolino, Rieke, Rieke, Rix, Robertson, Stark, Tacchella, Williams, Willmer, Baker, Baum, Bhatawdekar, Boyett, Chen, Chevallard, Circosta, Curti, Danhaive, DeCoursey, {de Graaff}, Dressler, Egami, Helton, Hviding, Ji, Jones, Kumari, L{\"u}tzgendorf, Laseter, Looser, Lyu, Maseda, Nelson, Parlanti, Perna, Pusk{\'a}s, Rawle, Rodr{\'i}guez Del~Pino, Sandles, Saxena, Scholtz, Sharpe, Shivaei, Silcock, Simmonds, Skarbinski, Smit, Stone, Suess, Sun, Tang, Topping, {\"U}bler, Villanueva, Wallace, Whitler, Witstok, \& Woodrum}]{eisensteinOverviewJWSTAdvanced2024} Eisenstein, D.~J., Willott, C., Alberts, S., {et~al.} 2025, \apjs, submitted (arXiv:2306.02465)

\bibitem[Finkelstein et al.(2025)]{Finkelstein2025} Finkelstein, S.~L., Bagley, M.~B., Arrabal Haro, P., et al.\ 2025, \apjl, 983, 1, L4

\bibitem[{Finkelstein {et~al.}(2023)Finkelstein, Bagley, Ferguson, Wilkins, Kartaltepe, Papovich, Yung, Haro, Behroozi, Dickinson, Kocevski, Koekemoer, Larson, Bail, Morales, {Perez-Gonzalez}, Burgarella, Dave, Hirschmann, Somerville, Wuyts, Bromm, Casey, Fontana, Fujimoto, Gardner, Giavalisco, Grazian, Grogin, Hathi, Hutchison, Jha, Jogee, Kewley, Kirkpatrick, Long, Lotz, Pentericci, Pierel, Pirzkal, Ravindranath, Ryan~Jr, Trump, Yang, Bhatawdekar, Bisigello, Buat, Calabro, Castellano, Cleri, Cooper, Croton, Daddi, Dekel, Elbaz, Franco, Gawiser, Holwerda, {Huertas-Company}, Jaskot, Leung, Lucas, Mobasher, Pandya, Tacchella, Weiner, \& Zavala}]{finkelsteinCEERSKeyPaper2023} Finkelstein, S.~L., Bagley, M.~B., Ferguson, H.~C., {et~al.} 2023, \apjl, 946, L13

\bibitem[{Frankel {et~al.}(2019)Frankel, Sanders, Rix, Ting, \& Ness}]{frankelInsideoutGrowthGalactic2019} Frankel, N., Sanders, J., Rix, H.-W., Ting, Y.-S., \& Ness, M. 2019, \apj, 884, 99

\bibitem[Fudamoto et al.(2020)]{fudamoto2020} Fudamoto, Y., Oesch, P.~A., Faisst, A., et al.\ 2020, \aap, 643, A4

\bibitem[Gardner et al.(2006)]{Gardner2006} Gardner, J.~P., Mather, J.~C., Clampin, M., et al.\ 2006, \ssr, 123, 4, 485

\bibitem[{{Gim{\'e}nez-Arteaga} {et~al.}(2023){Gim{\'e}nez-Arteaga}, Oesch, Brammer, Valentino, Mason, Weibel, Barrufet, Fujimoto, Heintz, Nelson, Strait, Suess, \& Gibson}]{gimenez-arteagaSpatiallyResolvedProperties2023a} {Gim{\'e}nez-Arteaga}, C., Oesch, P.~A., Brammer, G.~B., {et~al.} 2023, \apj, 948, 126

\bibitem[{Graham \& Worley(2008)}]{grahamInclinationDustcorrectedGalaxy2008} Graham, A.~W., \& Worley, C.~C. 2008, \mnras, 388, 1708

\bibitem[Harikane et al.(2025)]{harikane2025} Harikane, Y., Inoue, A.~K., Ellis, R.~S., et al.\ 2025, \apj, 980, 1, 138

\bibitem[Harikane et al.(2023)]{harikaneJWSTNIRSpecFirst2023} Harikane, Y., Zhang, Y., Nakajima, K., et al.\ 2023, \apj, 959, 39

\bibitem[{Harris {et~al.}(2020)Harris, Millman, {van der Walt}, Gommers, Virtanen, Cournapeau, Wieser, Taylor, Berg, Smith, Kern, Picus, Hoyer, {van Kerkwijk}, Brett, Haldane, {del R{\'i}o}, Wiebe, Peterson, {G{\'e}rard-Marchant}, Sheppard, Reddy, Weckesser, Abbasi, Gohlke, \& Oliphant}]{harrisArrayProgrammingNumPy2020} Harris, C.~R., Millman, K.~J., {van der Walt}, S.~J., {et~al.} 2020, \nat, 585, 357

\bibitem[Harvey et al.(2025)]{Harvey2025} Harvey, T., Conselice, C.~J., Adams, N.~J., et al.\ 2025, \apj, 978, 1, 89

\bibitem[Hasheminia et al.(2024)]{Hasheminia2024} Hasheminia, M., Mosleh, M., Hosseini-ShahiSavandi, S. Z., \& Tacchella, S. 2024, \apj, 975, 252

\bibitem[{H{\"a}u{\ss}ler {et~al.}(2013)H{\"a}u{\ss}ler, Bamford, Vika, Rojas, Barden, Kelvin, Alpaslan, Robotham, Driver, Baldry, Brough, Hopkins, Liske, Nichol, Popescu, \& Tuffs}]{hausslerMegaMorphMultiwavelengthMeasurement2013} H{\"a}u{\ss}ler, B., Bamford, S.~P., Vika, M., {et~al.} 2013, \mnras, 430, 330

\bibitem[{H{\"a}u{\ss}ler {et~al.}(2022)H{\"a}u{\ss}ler, Vika, Bamford, Johnston, Brough, Casura, Holwerda, Kelvin, \& Popescu}]{hausslerGALAPAGOS2GalfitMGAMA2022} H{\"a}u{\ss}ler, B., Vika, M., Bamford, S.~P., {et~al.} 2022, \aap, 664, A92

\bibitem[{Ho {et~al.}(2011)Ho, Li, Barth, Seigar, \& Peng}]{hoCarnegieIrvineGalaxySurvey2011} Ho, L.~C., Li, Z.-Y., Barth, A.~J., Seigar, M.~S., \& Peng, C.~Y. 2011, \apjs, 197, 21

\bibitem[Hopkins et al.(2010)]{2010MNRAS.401.1099H} Hopkins, P.~F., Bundy, K., Hernquist, L., et al.\ 2010, \mnras, 401, 1099

\bibitem[{Huang {et~al.}(2013)Huang, Ho, Peng, Li, \& Barth}]{huangCarnegieIrvineGalaxySurvey2013} Huang, S., Ho, L.~C., Peng, C.~Y., Li, Z.-Y., \& Barth, A.~J. 2013, \apj, 766, 47

\bibitem[{Hunter(2007)}]{hunterMatplotlib2DGraphics2007} Hunter, J.~D. 2007, Computing in Science \& Engineering, 9, 90

\bibitem[Ji \& Giavalisco(2023)]{Ji2023} Ji, Z. \& Giavalisco, M.\ 2023, \apj, 943, 1, 54

\bibitem[{Ji {et~al.}(2025)Ji, Williams, Suess, Tacchella, Johnson, Robertson, Alberts, Baker, Baum, Bhatawdekar, Bonaventura, Boyett, Bunker, Carniani, Charlot, Chen, Chevallard, {Curtis-Lake}, D'Eugenio, {de Graaff}, DeCoursey, Egami, Eisenstein, Hainline, Hausen, Helton, Looser, Lyu, Maiolino, Maseda, Nelson, Rieke, Rieke, Rix, Sandles, Sun, {\"U}bler, Willmer, Willott, \& Witstok}]{jiJADESRestframeUVtoNIR2024} Ji, Z., Williams, C.~C., Suess, K.~A., {et~al.} 2025, \apj, submitted (arXiv:2401.00934)

\bibitem[Jia et al.(2024)]{JiaSizeGrowth2024} Jia, C., Wang, E., Wang, H., et al.\ 2024, \apj, 977, 2, 165

\bibitem[{Kelvin {et~al.}(2012)Kelvin, Driver, Robotham, Hill, Alpaslan, Baldry, Bamford, {Bland-Hawthorn}, Brough, Graham, H{\"a}ussler, Hopkins, Liske, Loveday, Norberg, Phillipps, Popescu, Prescott, Taylor, \& Tuffs}]{kelvinGalaxyMassAssembly2012} Kelvin, L.~S., Driver, S.~P., Robotham, A. S.~G., {et~al.} 2012, \mnras, 421, 1007

\bibitem[{Kennedy {et~al.}(2015)Kennedy, Bamford, Baldry, H{\"a}u{\ss}ler, Holwerda, Hopkins, Kelvin, Lange, Moffett, Popescu, Taylor, Tuffs, Vika, \& Vulcani}]{kennedyGalaxyMassAssembly2015} Kennedy, R., Bamford, S.~P., Baldry, I., {et~al.} 2015, \mnras, 454, 806

\bibitem[Kocevski et al.(2023)]{Kocevski+2023+AGN} Kocevski, D.~D., Barro, G., McGrath, E.~J., et al.\ 2023, \apjl, 946, 1, L14

\bibitem[{Koekemoer {et~al.}(2011)Koekemoer, Faber, Ferguson, Grogin, Kocevski,
  Koo, Lai, Lotz, Lucas, McGrath, Ogaz, Rajan, Riess, Rodney, Strolger,
  Casertano, Castellano, Dahlen, Dickinson, Dolch, Fontana, Giavalisco,
  Grazian, Guo, Hathi, Huang, {van der Wel}, Yan, Acquaviva, Almaini, Ashby,
  Barden, Bell, Bournaud, Brown, Caputi, Cassata, Challis, Chary, Cheung,
  Cirasuolo, Conselice, Cooray, Croton, Daddi, Dav{\'e}, {de Mello}, {de
  Ravel}, Dekel, Donley, Dunlop, Dutton, Elbaz, Fazio, Filippenko, Finkelstein,
  Frazer, Gardner, Garnavich, Gawiser, Gruetzbauch, Hartley, H{\"a}ussler,
  Herrington, Hopkins, Huang, Jha, Johnson, Kartaltepe, Khostovan, Kirshner,
  Lani, Lee, Li, Madau, McCarthy, McIntosh, McLure, McPartland, Mobasher,
  Moreira, Mortlock, Moustakas, Mozena, Nandra, Newman, Nielsen, Niemi, Noeske,
  Papovich, Pentericci, Pope, Primack, Ravindranath, Reddy, Renzini, Rix,
  Robaina, Rosario, Rosati, Salimbeni, Scarlata, Siana, Simard, Smidt, Snyder,
  Somerville, Spinrad, Straughn, Telford, Teplitz, Trump, Vargas, Villforth,
  Wagner, Wandro, Wechsler, Weiner, Wiklind, Wild, Wilson, Wuyts, \&
  Yun}]{koekemoerCANDELSCosmicAssembly2011}
Koekemoer, A.~M., Faber, S.~M., Ferguson, H.~C., {et~al.} 2011, \apjs, 197, 36,

\bibitem[Kokorev et al.(2024)]{KokorevLRD2024ApJ} Kokorev, V., Caputi, K.~I., Greene, J.~E., et al.\ 2024, \apj, 968, 38

\bibitem[{Kron(1980)}]{kronPhotometryCompleteSample1980} Kron, R.~G. 1980, \apjs, 43, 305

\bibitem[Labb\'e et al.(2025)]{Labbe2025} Labb\'e, I., Greene, J.~E., Bezanson, R., et al.\ 2025, \apj, 978, 1, 92

\bibitem[Lapiner et al.(2023)]{LapinerBN2023} Lapiner, S., Dekel, A., Freundlich, J., et al.\ 2023, \mnras, 522, 4515

\bibitem[{Larson {et~al.}(2023)Larson, Hutchison, Bagley, Finkelstein, Yung, Somerville, Hirschmann, Brammer, Holwerda, Papovich, Morales, \& Wilkins}]{larsonSpectralTemplatesOptimal2023} Larson, R.~L., Hutchison, T.~A., Bagley, M., {et~al.} 2023, \apj, 958, 141

\bibitem[Leja et al.(2019)]{Leja2019} Leja, J., Carnall, A.~C., Johnson, B.~D., et al.\ 2019, \apj, 876, 1, 3

\bibitem[{Li {et~al.}(2019)Li, Narayanan, \& Dav{\'e}}]{liDusttogasDusttometalRatio2019} Li, Q., Narayanan, D., \& Dav{\'e}, R. 2019, \mnras, 490, 1425

\bibitem[Li et al.(2023)]{Liyang+2023} Li, Y.~A., Ho, L.~C., Shangguan, J., et al.\ 2023, \apjs, 267, 17

\bibitem[Looser et al.(2025)]{Looser2025} Looser, T.~J., D'Eugenio, F., Maiolino, R., et al.\ 2025, \aap, 697, A88

\bibitem[Lyu et al.(2025)]{Lyu2025} Lyu, Y., Magnelli, B., Elbaz, D., et al.\ 2025, \aap, 693, A313

\bibitem[{Magnelli {et~al.}(2023)Magnelli, {G{\'o}mez-Guijarro}, Elbaz, Daddi, Papovich, Shen, Arrabal~Haro, Bagley, Bell, Buat, Costantin, Dickinson, Finkelstein, Gardner, {Jim{\'e}nez-Andrade}, Kartaltepe, Koekemoer, Lyu, {P{\'e}rez-Gonz{\'a}lez}, Pirzkal, Tacchella, {de la Vega}, Wuyts, Yang, Yung, \& Zavala}]{magnelliCEERSMIRIDeciphers2023} Magnelli, B., {G{\'o}mez-Guijarro}, C., Elbaz, D., {et~al.} 2023, \aap, 678, A83

\bibitem[{Marian {et~al.}(2018)Marian, Ziegler, Kuchner, \& Verdugo}]{marianColorGradientsReflect2018} Marian, V., Ziegler, B., Kuchner, U., \& Verdugo, M. 2018, \aap, 617, A34

\bibitem[{Matharu {et~al.}(2023)Matharu, Muzzin, Sarrouh, Brammer, Abraham, Asada, Brada{\v c}, Desprez, Martis, Mowla, Noirot, Sawicki, Strait, Willott, Gould, Grindlay, \& Harshan}]{matharuFirstLookSpatially2023} Matharu, J., Muzzin, A., Sarrouh, G., {et~al.} 2023, \apjl, 949, L11

\bibitem[Matharu et al.(2024)]{matharuFirstLookSpatially2024} Matharu, J., Nelson, E.~J., Brammer, G., et al.\ 2024, \aap, 690, A64

\bibitem[{Miller {et~al.}(2023)Miller, {van Dokkum}, \& Mowla}]{millerColorGradientsHalfmass2023a} Miller, T.~B., {van Dokkum}, P., \& Mowla, L. 2023, \apj, 945, 155

\bibitem[{Miller {et~al.}(2022)Miller, Whitaker, Nelson, {van Dokkum}, Bezanson, Brammer, Heintz, Leja, Suess, \& Weaver}]{millerEarlyJWSTImaging2022} Miller, T.~B., Whitaker, K.~E., Nelson, E.~J., {et~al.} 2022, \apjl, 941, L37

\bibitem[{Mitsuhashi {et~al.}(2024)Mitsuhashi, Tadaki, Ikeda, {Herrera-Camus}, Aravena, De~Looze, Schreiber, {Gonz{\'a}lez-L{\'o}pez}, Spilker, Assef, Bouwens, {Barcos-Munoz}, Birkin, Bowler, Rivera, Davies, Da~Cunha, {D{\'i}az-Santos}, Ferrara, Fisher, Lee, Li, Lutz, Rela{\~n}o, Naab, Palla, Posses, Solimano, Tacconi, {\"U}bler, {van der Giessen}, \& Veilleux}]{mitsuhashiALMACRISTALSurveyWidespread2024} Mitsuhashi, I., Tadaki, K.-i., Ikeda, R., {et~al.} 2024, \aap, 690, A197

\bibitem[{Mo {et~al.}(1998)Mo, Mao, \& White}]{moFormationGalacticDisks1998} Mo, H.~J., Mao, S., \& White, S. D.~M. 1998, \mnras, 295, 319

\bibitem[{Morishita {et~al.}(2024)Morishita, Stiavelli, Chary, Trenti, Bergamini, Chiaberge, Leethochawalit, {Roberts-Borsani}, Shen, \& Treu}]{morishitaEnhancedSubkpcScale2024} Morishita, T., Stiavelli, M., Chary, R.-R., {et~al.} 2024, \apj, 963, 9

\bibitem[{Mosleh {et~al.}(2012)Mosleh, Williams, Franx, Gonzalez, Bouwens, Oesch, Labbe, Illingworth, \& Trenti}]{moslehEvolutionMassSizeRelation2012} Mosleh, M., Williams, R.~J., Franx, M., {et~al.} 2012, \apj, 756, L12

\bibitem[{{Mu{\~n}oz-Mateos} {et~al.}(2007){Mu{\~n}oz-Mateos}, {Gil de Paz}, Boissier, Zamorano, Jarrett, Gallego, \& Madore}]{munoz-mateosSpecificStarFormation2007} {Mu{\~n}oz-Mateos}, J.~C., {Gil de Paz}, A., Boissier, S., {et~al.} 2007, \apj, 658, 1006

\bibitem[{Nedkova {et~al.}(2024{\natexlab{a}})Nedkova, H{\"a}u{\ss}ler, Marchesini, Brammer, Feinstein, Johnston, Kartaltepe, Koekemoer, Martis, Muzzin, Rafelski, Shipley, Skelton, Stefanon, {van der Wel}, \& Whitaker}]{nedkovaBulgeDiscDecomposition2024} Nedkova, K.~V., H{\"a}u{\ss}ler, B., Marchesini, D., {et~al.} 2024{\natexlab{a}}, \mnras, 532, 3747

\bibitem[{Nedkova {et~al.}(2024{\natexlab{b}})Nedkova, Rafelski, Teplitz, Mehta, DeGroot, Ravindranath, Alavi, Beckett, Grogin, H{\"a}u{\ss}ler, Koekemoer, Oyarz{\'u}n, Prichard, Revalski, Snyder, Sunnquist, Wang, Windhorst, Chartab, Conselice, Guo, Hathi, Hayes, Ji, Kim, Lucas, Mobasher, O'Connell, Sattari, Smith, Taamoli, Yung, \& Team}]{nedkovaUVCANDELSRoleDust2024} Nedkova, K.~V., Rafelski, M., Teplitz, H.~I., {et~al.} 2024{\natexlab{b}}, \apj, 970, 188

\bibitem[Nelson et al.(2021)]{nelson2021MNRAS.508..219N} Nelson, E.~J., Tacchella, S., Diemer, B., et al.\ 2021, \mnras, 508, 219

\bibitem[{Nelson {et~al.}(2016)Nelson, {van Dokkum}, F{\"o}rster~Schreiber, Franx, Brammer, Momcheva, Wuyts, Whitaker, Skelton, Fumagalli, Hayward, Kriek, Labb{\'e}, Leja, Rix, Tacconi, {van der Wel}, {van den Bosch}, Oesch, Dickey, \& Ulf~Lange}]{nelsonWhereStarsForm2016} Nelson, E.~J., {van Dokkum}, P.~G., F{\"o}rster~Schreiber, N.~M., {et~al.} 2016, \apj, 828, 27

\bibitem[Oke \& Gunn(1983)]{Oke1983} Oke, J. B., \& Gunn, J. E. 1983, \apj, 266, 713

\bibitem[{Ono {et~al.}(2024)Ono, Harikane, Ouchi, Nakajima, Isobe, Shibuya, Nakane, Umeda, Xu, \& Zhang}]{onoCensusRestframeOptical2024} Ono, Y., Harikane, Y., Ouchi, M., et al.\ 2024, \pasj, 76, 219

\bibitem[Onoue et al.(2023)]{Onouemasafusa+2023+agn} Onoue, M., Inayoshi, K., Ding, X., et al.\ 2023, \apjl, 942, 1, L17

\bibitem[Pandya et al.(2024)]{pandyaGalaxiesGoingBananas2023} Pandya, V., Zhang, H., Huertas-Company, M., et al.\ 2024, \apj, 963, 54

\bibitem[Park et al.(2010)]{Park+2010+irselectedagn} Park, S.~Q., Barmby, P., Willner, S.~P., et al.\ 2010, \apj, 717, 2, 1181

\bibitem[{Peng {et~al.}(2002)Peng, Ho, Impey, \& Rix}]{pengDetailedStructuralDecomposition2002} Peng, C.~Y., Ho, L.~C., Impey, C.~D., \& Rix, H.-W. 2002, \aj, 124, 266

\bibitem[Peng et al.(2010)]{pengDETAILEDDECOMPOSITIONGALAXY2010} Peng, C.~Y., Ho, L.~C., Impey, C.~D., \& Rix, H.-W. 2010, \aj, 139, 2097

\bibitem[{P{\'e}rez {et~al.}(2013)P{\'e}rez, Cid~Fernandes, Gonz{\'a}lez~Delgado, {Garc{\'i}a-Benito}, S{\'a}nchez, Husemann, Mast, Rod{\'o}n, Kupko, Backsmann, {de Amorim}, {van de Ven}, Walcher, Wisotzki, {Cortijo-Ferrero}, \& {CALIFA Collaboration}}]{perezEvolutionGalaxiesResolved2013} P{\'e}rez, E., Cid~Fernandes, R., Gonz{\'a}lez~Delgado, R.~M., {et~al.} 2013, \apj, 764, L1

\bibitem[P{\'e}rez-Montero et al.(2016)]{Perez-Montero+2016} P{\'e}rez-Montero, E., Garc{\'\i}a-Benito, R., V{\'\i}lchez, J.~M., et al.\ 2016, \aap, 595, A62

\bibitem[{Pezzulli {et~al.}(2017)Pezzulli, Fraternali, \& Binney}]{pezzulliAngularMomentumCosmological2017} Pezzulli, G., Fraternali, F., \& Binney, J. 2017, \mnras, 467, 311

\bibitem[{{Planck Collaboration} {et~al.}(2020){Planck Collaboration}, Aghanim, Akrami, Ashdown, Aumont, Baccigalupi, Ballardini, Banday, Barreiro, Bartolo, Basak, Battye, Ben2024_home_1354abed, Bernard, Bersanelli, Bielewicz, Bock, Bond, Borrill, Bouchet, Boulanger, Bucher, Burigana, Butler, Calabrese, Cardoso, Carron, Challinor, Chiang, Chluba, Colombo, Combet, Contreras, Crill, Cuttaia, {de Bernardis}, {de Zotti}, Delabrouille, Delouis, Di~Valentino, Diego, Dor{\'e}, Douspis, Ducout, Dupac, Dusini, Efstathiou, Elsner, En{\ss}lin, Eriksen, Fantaye, Farhang, Fergusson, {Fernandez-Cobos}, Finelli, Forastieri, Frailis, Fraisse, Franceschi, Frolov, Galeotta, Galli, Ganga, {G{\'e}nova-Santos}, Gerbino, Ghosh, {Gonz{\'a}lez-Nuevo}, G{\'o}rski, Gratton, Gruppuso, Gudmundsson, Hamann, Handley, Hansen, Herranz, Hildebrandt, Hivon, Huang, Jaffe, Jones, Karakci, Keih{\"a}nen, Keskitalo, Kiiveri, Kim, Kisner, Knox, Krachmalnicoff, Kunz, {Kurki-Suonio}, Lagache, Lamarre, Lasenby, Lattanzi, Lawrence, Le~Jeune, Lemos, Lesgourgues, Levrier, Lewis, Liguori, Lilje, Lilley, Lindholm, {L{\'o}pez-Caniego}, Lubin, Ma, {Mac{\'i}as-P{\'e}rez}, Maggio, Maino, Mandolesi, Mangilli, {Marcos-Caballero}, Maris, Martin, Martinelli, {Mart{\'i}nez-Gonz{\'a}lez}, Matarrese, Mauri, McEwen, Meinhold, Melchiorri, Mennella, Migliaccio, Millea, Mitra, {Miville-Desch{\^e}nes}, Molinari, Montier, Morgante, Moss, Natoli, {N{\o}rgaard-Nielsen}, Pagano, Paoletti, Partridge, Patanchon, Peiris, Perrotta, Pettorino, Piacentini, Polastri, Polenta, Puget, Rachen, Reinecke, Remazeilles, Renzi, Rocha, Rosset, Roudier, {Rubi{\~n}o-Mart{\'i}n}, {Ruiz-Granados}, Salvati, Sandri, Savelainen, Scott, Shellard, Sirignano, Sirri, Spencer, Sunyaev, {Suur-Uski}, Tauber, Tavagnacco, Tenti, Toffolatti, Tomasi, Trombetti, Valenziano, Valiviita, Van~Tent, Vibert, Vielva, Villa, Vittorio, Wandelt, Wehus, White, White, Zacchei, \& Zonca}]{planckcollaborationPlanck2018Results2020} {Planck Collaboration}, Aghanim, N., Akrami, Y., {et~al.} 2020, \aap, 641, A6

\bibitem[{{Rieke} {et~al.}(2023){Rieke}, {Kelly}, {Misselt}, {Stansberry}, {Boyer}, {Beatty}, {Egami}, {Florian}, {Greene}, {Hainline}, {Leisenring}, {Roellig}, {Schlawin}, {Sun}, {Tinnin}, {Williams}, {Willmer}, {Wilson}, {Clark}, {Rohrbach}, {Brooks}, {Canipe}, {Correnti}, {DiFelice}, {Gennaro}, {Girard}, {Hartig}, {Hilbert}, {Koekemoer}, {Nikolov}, {Pirzkal}, {Rest}, {Robberto}, {Sunnquist}, {Telfer}, {Wu}, {Ferry}, {Lewis}, {Baum}, {Beichman}, {Doyon}, {Dressler}, {Eisenstein}, {Ferrarese}, {Hodapp}, {Horner}, {Jaffe}, {Johnstone}, {Krist}, {Martin}, {McCarthy}, {Meyer}, {Rieke}, {Trauger}, \& {Young}}]{2023PASP..135b8001R} {Rieke}, M.~J., {Kelly}, D.~M., {Misselt}, K., {et~al.} 2023, \pasp, 135, 028001

\bibitem[{Rigby {et~al.}(2023)Rigby, Perrin, McElwain, Kimble, Friedman, Lallo, Doyon, Feinberg, Ferruit, Glasse, Rieke, Rieke, Wright, Willott, Colon, Milam, Neff, Stark, Valenti, Abell, Abney, {Abul-Huda}, Acton, Adams, Adler, Aguilar, Ahmed, Albert, Alberts, Aldridge, Allen, Altenburg, {\'A}lvarez-M{\'a}rquez, de~Oliveira, Andersen, Anderson, Anderson, Argyriou, Armstrong, Arribas, Artigau, Arvai, Atkinson, Bacon, Bair, Banks, Barrientes, Barringer, Bartosik, Bast, Baudoz, Beatty, Bechtold, Beck, Bergeron, Bergkoetter, Bhatawdekar, Birkmann, Blazek, Blome, Boccaletti, B{\"o}ker, Boia, Bonaventura, Bond, Bosley, Boucarut, Bourque, Bouwman, Bower, Bowers, Boyer, Bradley, Brady, Braun, Breda, Bresnahan, Bright, Britt, Bromenschenkel, Brooks, Brooks, Brown, Brown, Brown, Bunker, Burger, Bushouse, Cale, Cameron, Cameron, Canipe, Caplinger, Caputo, Cara, Carey, Carniani, Carrasquilla, Carruthers, Case, Catherine, Chance, Chapman, Charlot, Charlow, Chayer, Chen, Cherinka, Chichester, Chilton, Chonis, Clampin, Clark, Clark, Coe, Coleman, Comber, Comeau, Connolly, Cooper, Cooper, Coppock, Correnti, Cossou, Coulais, Coyle, Cracraft, Curti, Cuturic, Davis, Davis, Dean, DeLisa, {deMeester}, Dencheva, Dencheva, DePasquale, Deschenes, Detre, Diaz, Dicken, DiFelice, Dillman, Dixon, Doggett, Donaldson, Douglas, DuPrie, Dupuis, Durning, Easmin, Eck, Edeani, Egami, Ehrenwinkler, Eisenhamer, Eisenhower, Elie, Elliott, Elliott, Ellis, Engesser, Espinoza, Etienne, Etxaluze, Falini, Feeney, Ferry, Filippazzo, Fincham, Fix, Flagey, Florian, Flynn, Fontanella, Ford, Forshay, Fox, Franz, Fu, Fullerton, Galkin, Galyer, Mar{\'i}n, Gardner, Gardner, Garland, Garrett, Gasman, Gaspar, Gaudreau, Gauthier, Geers, Geithner, Gennaro, Giardino, Girard, Giuliano, Glassmire, Glauser, Glazer, Godfrey, Golimowski, Gollnitz, Gong, Gonzaga, Gordon, Gordon, Goudfrooij, Greene, Greenhouse, Grimaldi, Groebner, Grundy, Guillard, Gutman, Ha, Haderlein, Hagedorn, Hainline, Haley, Hami, Hamilton, Hammel, Hansen, Harkins, Harr, Hart, Hart, Hartig, Hashimoto, Haskins, Hathaway, Havey, Hayden, Hecht, {Heller-Boyer}, Henriques, Henry, Hermann, Hernandez, Hesman, Hicks, Hilbert, Hines, Hoffman, Holfeltz, Holler, Hoppa, Hott, Howard, Howard, Hunter, Hunter, Hurst, Husemann, Hustak, Ignat, Illingworth, Irish, Jackson, Jahromi, Jakobsen, James, James, Januszewski, Jenkins, Jirdeh, Johnson, Johnson, Jones, Jones, Jones, Jones, Jordan, Jordan, Jurczyk, Jurling, Kaleida, Kalmanson, Kammerer, Kang, Kao, Karakla, Kavanagh, Kelly, Kendrew, Kennedy, Kenny, {Keski-kuha}, Keyes, Kidwell, Kinzel, Kirk, Kirkpatrick, Kirshenblat, Klaassen, Knapp, Knight, Knollenberg, Koehler, Koekemoer, Kovacs, Kulp, Kumari, Kyprianou, Massa, Labador, Labiano, Lagage, Lajoie, Lallo, Lam, Lamb, Lambros, Lampenfield, Langston, Larson, Law, Lawrence, Lee, Leisenring, Lepo, Leveille, Levenson, Levine, Levy, Lewis, Lewis, Libralato, Lightsey, Link, Liu, Lo, Lockwood, Logue, Long, Long, Loomis, {Lopez-Caniego}, Alvarez, {Love-Pruitt}, Lucy, Luetzgendorf, Maghami, Maiolino, Major, Malla, Malumuth, Manjavacas, Mannfolk, Marrione, Marston, Martel, Maschmann, Masci, Masciarelli, Maszkiewicz, Mather, McKenzie, McLean, McMaster, Melbourne, Mel{\'e}ndez, Menzel, Merz, Meyett, Meza, Miskey, Misselt, Moller, Morrison, Morse, Moseley, Mosier, Mountain, Mueckay, Mueller, Mullally, Murphy, Murray, Murray, Mustelier, Muzerolle, Mycroft, Myers, Myrick, Nanavati, Nance, Nayak, Naylor, Nelan, Nickson, Nielson, {Nieto-Santisteban}, Nikolov, {Noriega-Crespo}, O'Shaughnessy, O'Sullivan, Ochs, Ogle, Oleszczuk, Olmsted, Osborne, Ottens, Owens, Pacifici, Pagan, Page, Park, Parrish, Patapis, Paul, Pauly, Pavlovsky, Pedder, Peek, {Pena-Guerrero}, Penanen, Perez, Perna, Perriello, Phillips, Pietraszkiewicz, Pinaud, Pirzkal, Pitman, Piwowar, Platais, Player, Plesha, Pollizi, Polster, Pontoppidan, Porterfield, Proffitt, Pueyo, Pulliam, Quirt, Neira, Alarcon, Ramsay, Rapp, Rapp, Rauscher, Ravindranath, Rawle, Regan, Reichard, Reis, Ressler, Rest, Reynolds, Rhue, Richon, Rickman, Ridgaway, Ritchie, Rix, Robberto, Robinson, Robinson, Robinson, Rock, Rodriguez, Pino, Roellig, Rohrbach, Roman, Romelfanger, Rose, Roteliuk, Roth, Rothwell, Rowlands, Roy, Royer, Royle, Rui, Rumler, Runnels, Russ, Rustamkulov, Ryden, Ryer, Sabata, Sabatke, Sabbi, Samuelson, Sapp, Sappington, Sargent, Sauer, Scheithauer, Schlawin, Schlitz, Schmitz, Schneider, Schreiber, Schulze, Schwab, Scott, Sembach, Shanahan, Shaughnessy, Shaw, Shawger, Shay, Sheehan, Shen, Sherman, Shiao, Shih, Shivaei, Sienkiewicz, Sing, Sirianni, Sivaramakrishnan, Skipper, Sloan, Slocum, Slowinski, Smith, Smith, Smith, Smith, Snyder, Soh, Sohn, Soto, Spencer, Stallcup, Stansberry, Starr, Starr, Stewart, Stiavelli, Straughn, Strickland, Stys, Summers, Sun, Sunnquist, Swade, Swam, Swaters, Swoish, Taylor, Taylor, Plate, Tea, Teague, Telfer, Temim, Thatte, Thompson, Thompson, Thomson, Tikkanen, Tippet, Todd, Toolan, Tran, Trejo, Truong, Tsukamoto, Tustain, Tyra, Ubeda, Underwood, Uzzo, Campen, Vandal, Vandenbussche, Vila, Volk, Wahlgren, Waldman, Walker, Wander, Warfield, Warner, Wasiak, Watkins, Weaver, Weilert, Weiser, Weiss, Weissman, Welty, West, Wheate, Wheatley, Wheeler, White, Whiteaker, Whitehouse, Whiteleather, Whitman, Williams, Willmer, Willoughby, Wilson, Wirth, Wislowski, Wolf, Wolfe, Wolff, Workman, Wright, Wu, Wu, Wymer, Yates, Yeager, Yeates, Yerger, Yoon, Young, Yu, Zak, Zeidler, Zhou, Zielinski, Zincke, \& Zonak}]{rigbySciencePerformanceJWST2023} Rigby, J., Perrin, M., McElwain, M., {et~al.} 2023, \pasp, 135, 048001

\bibitem[{Ryder \& Dopita(1994)}]{ryderRelationshipPresentStar1994} Ryder, S.~D., \& Dopita, M.~A. 1994, \apj, 430, 142

\bibitem[Schlafly \& Finkbeiner (2011)]{Schlafly2011} Schlafly, E. F., \& Finkbeiner, D. P. 2011, \apj, 737, 103

\bibitem[{{S{\'e}rsic}(1968)}]{Sersic1968} S\'ersic, J. L. 1968, Atlas de Galaxias Australes (C\'ordoba: Obs. Astron., Univ. Nac. C\'ordoba)

\bibitem[{Shen {et~al.}(2003)Shen, Mo, White, Blanton, Kauffmann, Voges, Brinkmann, \& Csabai}]{shenSizeDistributionGalaxies2003} Shen, S., Mo, H.~J., White, S. D.~M., {et~al.} 2003, \mnras, 343, 978

\bibitem[{Shibuya {et~al.}(2015)Shibuya, Ouchi, \& Harikane}]{shibuyaMorphologies1900002015} Shibuya, T., Ouchi, M., \& Harikane, Y. 2015, \apjs, 219, 15

\bibitem[Simons et al.(2021)]{Simons+2021} Simons, R.~C., Papovich, C., Momcheva, I., et al.\ 2021, \apj, 923, 203

\bibitem[Skelton et al.(2014)]{Skelton2014ApJS} Skelton, R.~E., Whitaker, K.~E., Momcheva, I.~G., et al.\ 2014, \apjs, 214, 24

\bibitem[Stark et al.(2024)]{StarkBreakBRD2024} Stark, D.~V., Tuttle, S., Tonnesen, S., et al.\ 2024, \apj, 971, 116

\bibitem[{Stefanon {et~al.}(2017)Stefanon, Yan, Mobasher, Barro, Donley, Fontana, Hemmati, Koekemoer, Lee, Lee, Nayyeri, Peth, Pforr, Salvato, Wiklind, Wuyts, Ashby, Castellano, Conselice, Cooper, Cooray, Dolch, Ferguson, Galametz, Giavalisco, Guo, Willner, Dickinson, Faber, Fazio, Gardner, Gawiser, Grazian, Grogin, Kocevski, Koo, Lee, Lucas, McGrath, Nandra, Newman, \& {van der Wel}}]{stefanonCANDELSMultiwavelengthCatalogs2017} Stefanon, M., Yan, H., Mobasher, B., {et~al.} 2017, \apjs, 229, 32

\bibitem[Steinmetz \& Navarro(2002)]{steinmetz2002a} Steinmetz, M. \& Navarro, J.~F.\ 2002, \na, 7, 155

\bibitem[{Suess {et~al.}(2019)Suess, Kriek, Price, \& Barro}]{suessHalfmassRadii0002019} Suess, K.~A., Kriek, M., Price, S.~H., \& Barro, G. 2019, \apj, 877, 103

\bibitem[Suess et al.(2021)]{Suess2021} Suess, K.~A., Kriek, M., Price, S.~H., et al.\ 2021, \apj, 915, 2, 87

\bibitem[Sun et al.(2024)]{sunStructureMorphologyGalaxies2023} Sun, W., Ho, L.~C., Zhuang, M.-Y., et al.\ 2024, \apj, 960, 104

\bibitem[Sun et al.(2025)]{SunXunda+2025} Sun, X., Wang, X., Ma, X., et al.\ 2025, \apj, 986, 2, 179


\bibitem[Tacchella et al.(2018)]{tacchella2018ApJ...859...56T} Tacchella, S., Carollo, C.~M., F{\"o}rster Schreiber, N.~M., et al.\ 2018, \apj, 859, 56

\bibitem[{Tacchella {et~al.}(2015)Tacchella, Carollo, Renzini, F{\"o}rster~Schreiber, Lang, Wuyts, Cresci, Dekel, Genzel, Lilly, Mancini, Newman, Onodera, Shapley, Tacconi, Woo, \& Zamorani}]{tacchellaEvidenceMatureBulges2015} Tacchella, S., Carollo, C.~M., Renzini, A., {et~al.} 2015, Science, 348, 314

\bibitem[Tacchella et al.(2016)]{Tacchella2016} Tacchella, S., Dekel, A., Carollo, C.~M., et al.\ 2016, \mnras, 457, 2790

\bibitem[Tacchella et al.(2020)]{Tacchella2020} Tacchella, S., Forbes, J.~C., \& Caplar, N.\ 2020, \mnras, 497, 1, 698

\bibitem[{{The Astropy Collaboration} {et~al.}(2013){The Astropy Collaboration}, Robitaille, Tollerud, Greenfield, Droettboom, Bray, Aldcroft, Davis, Ginsburg, {Price-Whelan}, Kerzendorf, Conley, Crighton, Barbary, Muna, Ferguson, Grollier, Parikh, Nair, G{\"u}nther, Deil, Woillez, Conseil, Kramer, Turner, Singer, Fox, Weaver, Zabalza, Edwards, Azalee~Bostroem, Burke, Casey, Crawford, Dencheva, Ely, Jenness, Labrie, Lim, Pierfederici, Pontzen, Ptak, Refsdal, Servillat, \& Streicher}]{theastropycollaborationAstropyCommunityPython2013} {The Astropy Collaboration}, Robitaille, T.~P., Tollerud, E.~J., {et~al.} 2013, \aap, 558, A33

\bibitem[Tissera et al.(2022)]{Tissera+2022} Tissera, P.~B., Rosas-Guevara, Y., Sillero, E., et al.\ 2022, \mnras, 511, 1667

\bibitem[Tomassetti et al.(2016)]{TomassettiProlate2016MNRAS} Tomassetti, M., Dekel, A., Mandelker, N., et al.\ 2016, \mnras, 458, 4477

\bibitem[Tonnesen et al.(2023)]{TonnesenCGM2023} Tonnesen, S., DeFelippis, D., \& Tuttle, S.\ 2023, \apj, 951, 16

\bibitem[{Treu {et~al.}(2023)Treu, Calabro, Castellano, Leethochawalit, Merlin, Fontana, Yang, Morishita, Trenti, Dressler, Mason, Paris, Pentericci, {Roberts-Borsani}, Vulcani, Boyett, Bradac, Glazebrook, Jones, Marchesini, Mascia, Nanayakkara, Santini, Strait, Vanzella, \& Wang}]{treuEarlyResultsGLASSJWST2023} Treu, T., Calabro, A., Castellano, M., {et~al.} 2023, \apjl, 942, L28

\bibitem[Tripodi et al.(2024)]{tripodiSpatiallyResolvedEmission2024} Tripodi, R., D'Eugenio, F., Maiolino, R., et al.\ 2024, \aap, 692, A184

\bibitem[Tuttle \& Tonnesen(2020)]{tuttleBreakBRD2020} Tuttle, S.~E., \& Tonnesen, S.\ 2020, \apj, 889, 188

\bibitem[{Vader {et~al.}(1988)Vader, Vigroux, {Lachieze-Rey}, \& Souviron}]{vaderThreeColorSurface1988} Vader, J.~P., Vigroux, L., {Lachieze-Rey}, M., \& Souviron, J. 1988, \aap, 203, 217

\bibitem[van der Wel et al.(2014{\natexlab{a}})]{vanderwelProlate2014} van der Wel, A., Chang, Y.-Y., Bell, E.~F., et al.\ 2014{\natexlab{a}}, \apjl, 792, L6

\bibitem[{{van der Wel} {et~al.}(2014{\natexlab{b}}){van der Wel}, Franx, {van Dokkum}, Skelton, Momcheva, Whitaker, Brammer, Bell, Rix, Wuyts, Ferguson, Holden, Barro, Koekemoer, Chang, McGrath, Haussler, Dekel, Behroozi, Fumagalli, Leja, Lundgren, Maseda, Nelson, Wake, Patel, Labbe, Faber, Grogin, \& Kocevski}]{vanderwel3DHSTCANDELSEvolution2014} {van der Wel}, A., Franx, M., {van Dokkum}, P.~G., {et~al.} 2014{\natexlab{b}}, \apj, 788, 28

\bibitem[{{van der Wel} {et~al.}(2024){van der Wel}, Martorano, Haussler, Nedkova, Miller, Brammer, {van de Ven}, Leja, Bezanson, Muzzin, Marchesini, {de Graaff}, Kriek, Bell, \& Franx}]{vanderwelStellarHalfMassRadii2024} {van der Wel}, A., Martorano, M., H{\"a}u{\ss}ler, B., {et~al.} 2024, \apj, 960, 53

\bibitem[van Dokkum et al.(2013)]{vandokkum2013ApJ...771L..35V} van Dokkum, P.~G., Leja, J., Nelson, E.~J., et al.\ 2013, \apjl, 771, L35

\bibitem[{{van Dokkum} {et~al.}(2010){van Dokkum}, Whitaker, Brammer, Franx, Kriek, Labbe, Marchesini, Quadri, Bezanson, Illingworth, Muzzin, Rudnick, Tal, \& Wake}]{vandokkumGrowthMassiveGalaxies2010} {van Dokkum}, P.~G., Whitaker, K.~E., Brammer, G., {et~al.} 2010, \apj, 709, 1018

\bibitem[{Vika {et~al.}(2013)Vika, Bamford, Haeussler, Rojas, Borch, \& Nichol}]{vikaMegaMorphMultiwavelengthMeasurement2013}
Vika, M., Bamford, S.~P., Haeussler, B., {et~al.} 2013, \mnras, 435, 623

\bibitem[{Vika {et~al.}(2015)Vika, Vulcani, Bamford, H{\"a}u{\ss}ler, \& Rojas}]{vikaMegaMorphClassifyingGalaxy2015} Vika, M., Vulcani, B., Bamford, S.~P., H{\"a}u{\ss}ler, B., \& Rojas, A.~L.  2015, \aap, 577, A97

\bibitem[{Virtanen {et~al.}(2020)Virtanen, Gommers, Oliphant, Haberland, Reddy, Cournapeau, Burovski, Peterson, Weckesser, Bright, {van der Walt}, Brett, Wilson, Millman, Mayorov, Nelson, Jones, Kern, Larson, Carey, Polat, Feng, Moore, VanderPlas, Laxalde, Perktold, Cimrman, Henriksen, Quintero, Harris, Archibald, Ribeiro, Pedregosa, \& {van Mulbregt}}]{virtanenSciPyFundamentalAlgorithms2020} Virtanen, P., Gommers, R., Oliphant, T.~E., {et~al.} 2020, Nature Methods, 17, 261

\bibitem[{Vulcani {et~al.}(2014)Vulcani, Bamford, H{\"a}u{\ss}ler, Vika, Rojas, Agius, Baldry, Bauer, Brown, Driver, Graham, Kelvin, Liske, Loveday, Popescu, Robotham, \& Tuffs}]{vulcaniGalaxyMassAssembly2014b} Vulcani, B., Bamford, S.~P., H{\"a}u{\ss}ler, B., {et~al.} 2014, \mnras, 441, 1340

\bibitem[{Wang {et~al.}(2024)Wang, Leja, Labb{\'e}, Bezanson, Whitaker, Brammer, Furtak, Weaver, Price, Zitrin, Atek, Coe, Cutler, Dayal, {van Dokkum}, Feldmann, Marchesini, Franx, Schreiber, Fujimoto, Geha, Glazebrook, {de Graaff}, Greene, Juneau, Kassin, Kriek, Khullar, Maseda, Mowla, Muzzin, Nanayakkara, Nelson, Oesch, Pacifici, Pan, Papovich, Setton, Shapley, Smit, Stefanon, Suess, Taylor, \& Williams}]{wangUNCOVERSurveyFirstLook2024} Wang, B., Leja, J., Labb{\'e}, I., {et~al.} 2024, \apjs, 270, 12

\bibitem[Wang et al.(2017)]{Wang2017MNRAS} Wang, W., Faber, S.~M., Liu, F.~S., et al.\ 2017, \mnras, 469, 4063

\bibitem[{Ward {et~al.}(2024)Ward, {de la Vega}, Mobasher, McGrath, Iyer, Calabro, Costantin, Dickinson, Holwerda, {Huertas-Company}, Hirschmann, Lucas, Pandya, Wilkins, Yung, Haro, Bagley, Finkelstein, Kartaltepe, Koekemoer, Papovich, \& Pirzkal}]{wardEvolutionSizeMassRelation2024} Ward, E.~M., {de la Vega}, A., Mobasher, B., {et~al.} 2024, \apj, 962, 176

\bibitem[Weaver et al.(2024)]{Weaver2024} Weaver, J.~R., Cutler, S.~E., Pan, R., et al.\ 2024, \apjs, 270, 1, 7

\bibitem[{White \& Frenk(1991)}]{whiteGalaxyFormationHierarchical1991} White, S. D.~M., \& Frenk, C.~S. 1991, \apj, 379, 52

\bibitem[{Williams {et~al.}(2009)Williams, Quadri, Franx, {van Dokkum}, \& Labb{\'e}}]{williamsDetectionQuiescentGalaxies2009a} Williams, R.~J., Quadri, R.~F., Franx, M., {van Dokkum}, P., \& Labb{\'e}, I.  2009, \apj, 691, 1879

\bibitem[{Yang {et~al.}(2022)Yang, Morishita, Leethochawalit, Castellano, Calabr{\`o}, Treu, Bonchi, Fontana, Mason, Merlin, Paris, Trenti, {Roberts-Borsani}, Bradac, Vanzella, Vulcani, Marchesini, Ding, Nanayakkara, Birrer, Glazebrook, Jones, Boyett, Santini, Strait, \& Wang}]{yangEarlyResultsGLASSJWST2022} Yang, L., Morishita, T., Leethochawalit, N., {et~al.} 2022, \apjl, 938, L17

\bibitem[Zaritsky et al.(1994)]{Zaritsky+1994} Zaritsky, D., Kennicutt, R.~C., \& Huchra, J.~P.\ 1994, \apj, 420, 87

\bibitem[Zhang et al.(2019)]{zhang3DIntrinsicShapes2019} Zhang, H., Primack, J.~R., Faber, S.~M., et al.\ 2019, \mnras, 484, 5170

\bibitem[{Zhang {et~al.}(2023)Zhang, Wuyts, Cutler, Mowla, Brammer, Momcheva, Whitaker, {van Dokkum}, Schreiber, Nelson, Schady, Villforth, Wake, \& {van der Wel}}]{zhangDustAttenuationDust2023} Zhang, J., Wuyts, S., Cutler, S.~E., {et~al.} 2023, \mnras, 524, 4128

\bibitem[{Zhuang \& Ho(2022)}]{zhuangStarformingMainSequence2022} Zhuang, M.-Y., \& Ho, L.~C. 2022, \apj, 934, 130

\bibitem[Zhuang et al.(2024)]{zhuangAGNsHostGalaxies2024} Zhuang, M.-Y., Li, J., \& Shen, Y.\ 2024, \apj, 962, 93

\bibitem[Zhuang \& Shen(2024)]{zhuangCharacterizationJWSTNIRCam2024} Zhuang, M.-Y., \& Shen, Y.\ 2024, \apj, 962, 139

\bibitem[{Zolotov {et~al.}(2015)Zolotov, Dekel, Mandelker, Tweed, Inoue, DeGraf, Ceverino, Primack, Barro, \& Faber}]{zolotovCompactionQuenchingHighz2015} Zolotov, A., Dekel, A., Mandelker, N., {et~al.} 2015, \mnras, 450, 2327

\end{thebibliography}
\end{document}